\documentclass[sn-mathphys,Numbered]{sn-jnl}

\usepackage{graphicx}%
\usepackage{multirow}%
\usepackage{amsmath,amssymb,amsfonts}%
\usepackage{amsthm}%
\usepackage{mathrsfs}%
\usepackage[title]{appendix}%
\usepackage{xcolor}%
\usepackage{textcomp}%
\usepackage{manyfoot}%
\usepackage{booktabs}%
\usepackage{algorithm}%
\usepackage{algorithmicx}%
\usepackage{algpseudocode}%
\usepackage{listings}%
\usepackage{graphicx} 
\usepackage{multirow}
\usepackage{tabularx}
\usepackage{hhline} 
\UseRawInputEncoding 
\renewenvironment{table}[1][]%
{\tableorg[#1]%
\tablebodyfont%
\renewcommand\footnotetext[2][]{{\removelastskip\vskip3pt%
\let\tablebodyfont\tablefootnotefont%
\hskip0pt\if!##1!\else{\smash{$^{##1}$}}\fi##2\par}}%
}{\endtableorg}
\usepackage{float}

\usepackage{url}
\usepackage{ulem}

\theoremstyle{thmstyleone}%
\theoremstyle{thmstyletwo}%

\theoremstyle{thmstylethree}%

\begin{document}

\title[Article Title]{How to Design and Deliver Courses for Higher Education in the AI Era: Insights from Exam Data Analysis}


\author*[1]{\fnm{Ahmad Samer} \sur{Wazan}}\email{ahmad.wazan@zu.ac.ae}

\author[2]{\fnm{Imran} \sur{ Taj}}\email{MuhammadImran.Taj@zu.ac.ae}

\author[3]{\fnm{Abdulhadi} \sur{Shoufan}}\email{abdulhadi.shoufan@ku.ac.ae}

\author[4]{\fnm{Romain} \sur{Laborde}}\email{laborde@irit.fr}

\author[5]{\fnm{Rémi} \sur{Venant}}\email{Remi.Venant@univ-lemans.fr}

\affil*[1]{\orgdiv{College of Technological Innovation}, \orgname{Zayed University}, \country{UAE}}

\affil[2]{\orgdiv{College of Interdisciplinary Studies}, \orgname{Zayed University}, \country{UAE}}

\affil[3]{\orgdiv{Department of EECS}, \orgname{Khalifa University}, \country{UAE}}

\affil[4]{\orgdiv{IRIT Laboratory}, \orgname{University Toulouse 3 Paul Sabatier}, \country{France}}

\affil[5]{\orgname{LeMans University}, \country{France}}


\abstract{In this position paper, we advocate for the idea that courses and exams in the AI era have to be designed based on two factors: (1) the strengths and limitations of AI, and (2) the pedagogical educational objectives. Based on insights from the Delors report on education \cite{b5}, we first address the role of education and recall the main objectives that educational institutes must strive to achieve independently of any technology. We then explore the strengths and limitations of AI, based on current advances in AI. We explain how courses and exams can be designed based on these strengths and limitations of AI, providing different examples in the IT, English, and Art domains. We show how we adopted a pedagogical approach that is inspired from the Socratic teaching method from January 2023 to May 2023. Then, we present the data analysis results of seven ChatGPT-authorized exams conducted between December 2022 and March 2023. Our exam data results show that there is no correlation between students' grades and whether or not they use ChatGPT to answer their exam questions. Finally, we present a new exam system that allows us to apply our pedagogical approach in the AI era.}




\maketitle

\section{Introduction}\label{sec1}

Recent advancements in Generative AI have impacted the education sector tremendously. A number of concerns have been raised regarding the strategies for adopting these AI tools, particularly Large Language Models (LLMs), in a way that facilitates the role of educators in maintaining the integrity of the learning process. 

However, integrating AI into education is not a novel idea. AI based assistance for educational purposes has been an active research area since the 1970s, and has resulted in successful integration of various technology tools in the education sector. The rise in computing power and big data analysis \cite{b35} over the last decade has laid the foundation for sophisticated deep learning algorithms with billions of fine-tuned parameters. Generative Pre-trained Transformer (GPT) model - the engine of ChatGPT - exploits the same deep learning mechanisms to draw its knowledge from massive data sets, thereby achieving the capability of generating swift responses acquired by reinforcement and supervised learning techniques. The capability to generate rapid responses while solving math problems, coding programming languages, or writing assignments, including college essays, has the massive potential to improve student learning outcomes. Their 24/7 availability as personalized learning assistants makes these chatbots a strong candidate for designing efficient pedagogical approaches that could be integrated into the education sector. 

Since the advent of ChatGPT, its integration implications into educational assessments have been a hot research topic. The initial buzz about ChatGPT has mostly focused on its strengths and limitations for tests and exams across various educational specialties - college level chemistry \cite{b36}, mathematics \cite{b37}, journalism \& media education \cite{b38}, parasitology \cite{b39}, pharmacology \cite{b40}, medical license examinations \cite{b41}, law exams \cite{b42}, finance research \cite{b43}, and business administration \cite{b44} - to name a few. These efforts demonstrate the various inconsistent levels of ChatGPT’s performance; both good \cite{b44} and bad \cite{b36} and in between. 

Rudolph \cite{b45} has explored the relevance of ChatGPT for higher education, and offered general recommendations for lecturers, such as using text generator detection software, avoiding formulaic assignments, asking students to write about recent events not available in the training data and analyse images and videos.  While there has been other similar short term solutions \cite{b820,b830}, we could not come across any effort dedicated towards \textbf{\textit{courses design}} nor \textbf{\textit{courses delivery}} in the AI era. Our work addresses this gap by studying the key factors that enable the design of courses in the AI era, and by focusing on the development and testing of a pedagogical approach to deliver courses in a way that allows students to enhance their human intelligence. 

To design courses in the AI era, educators have to continuously observe the strengths and limitations of AI and ML. The strengths of AI allow educators to integrate AI tools into their courses to provide personalized education to their students. On the other hand, the limitations of AI allow educators to design activities and assessments that students cannot solve with a simple click of buttons. Being aware of AI's limitations is more difficult than understanding its strengths, so a significant part of this paper is focused on presenting the limitations of AI in an abstract manner.

In addition, designing courses and exams in the AI era must consider the level of education, the learning outcomes as well as the current problems that education faces. Therefore, we define the role of education based on the Delors report in a dedicated section, which is  independent of any technology to be integrated in education. Then, we highlight some of the current ineffective practices in the higher education system, such as the excessive use of memory-based exams and the over-reliance on textbooks. 

In terms of course delivery, we developed a pedagogical approach inspired by the Socratic method \cite{b32}, which stimulates the critical thinking of students. In each class session, we posed open-ended questions to our students, encouraging them to engage in a dialogue with ChatGPT before answering. We then systematically discussed our students' answers and follow-up questions, explaining why they were unable to obtain specific answers. Some of the questions were designed to explore the limitations of AI and were inspired by the class discussions, while others could have been answered by LLMs, but only after a series of follow-up questions.

We adopted two strategies to ensure the validity of our pedagogical approach:
\begin{itemize}
    \item Analysing students answers to exams conducted between December 2022 and March 2023. Our data analysis of the exam results revealed no correlation between students' grades and whether or not they use ChatGPT to answer their exam questions. 
    \item Hearing the concerns of educators working in various fields and universities. Between January and June 2023, we presented our pedagogical approach to educators working in Math, IT, English Teaching, Human studies and Art at nine universities across five different countries (UAE, Turkey, Poland and USA). The feedback about these presentation allowed us to collect and discuss various concerns from educators working in various domains. Based on these discussions, we defined in section \ref{sec7} a set of activities in different fields that prove the relevance of our pedagogical approach. 
\end{itemize}

The rest of the paper is organized as follows. In section 2, we explore the different reactions of stakeholders, including the education sector, to ChatGPT. In section 3, we discuss the role of education and analyze the current barriers to learning in the higher education system. Section 4 highlights the limitations and strengths of using AI in the education sector and provides insights into how these identified strengths and limitations can be utilized to design courses and exams in the AI era. Section 5 examines the internal working principles of LLMs, like ChatGPT. In section 6, we delve into how the working principles of these LLMs can be strategically incorporated into course and exam design, creating a pedagogical approach aligned with the Socratic method, thereby effectively addressing the higher education barriers analyzed in section 3. In section 7, we present eight case studies conducted on actual university students, covering multiple subject areas, to validate the proposed pedagogical approach outlined in section 6. The data analysis in section 8 demonstrates that when the courses and exams are designed according to the proposed pedagogical approach, the usage of LLMs like ChatGPT has no impact on student performance. Finally, we give our conclusions in section 9. 

\section{The World's Reaction to ChatGPT}\label{sec2}

Since the release of ChatGPT in November 2022, it has received a mix of enthusiastic and pessimistic responses in various sectors, including education where opinions split between considering LLMs as positive or negative, and everything in between. 

The Economist \cite{b46} summarized the reactions by pointing out the ability of GPT3 to be eerily human-like – for better and for worse. Elon Musk, CEO of Tesla, described ChatGPT to be "scary [sic!] good", warning about the risks of Generative AI becoming uncontrollable, and has advocated for proactive measures to ensure it is not misused. Along with Steve Wozniak, the co-founder of Apple, and Emad Mostaque, the CEO of Stability AI, Musk signed a petition calling for an end to gigantic AI experimentation \cite{b47}. Geoffrey Hinton, deemed as the pioneer of back-propagation deep learning technique of neural networks \cite{b48} has emphasized that the risks associated with Generative AI require more immediate consideration than those associated with climate change \cite{b840}. Since his resignation from Google, he has spoken out against the organization acting irresponsibly in the AI development of Bard \cite{b49}, without putting in place the necessary safeguards and regulations. Hinton’s stance is endorsed by prominent AI ethics researchers such as Timnit Gebru \cite{b50} who has also raised the concerns about the lack of transparency. Many other well-known organizations such as Verizon, JPMorgan Chase, Apple, and Amazon have restricted the usage of ChatGPT by their employees, referring to data security concerns \cite{b860}. 

In an attempt to address concerns from such business giants, OpenAI has taken measures to give users more control over the data used to train ChatGPT. The education sector is vigilant when it comes to the capabilities of ChatGPT to generate human-like text, which was referred to as the industry’s next big disrupter for industry and that such models are getting terrifyingly good at writing \cite{b51}. The Telegraph stated that “we should be very worried about AI-generated text” \cite{b52}. Noam Chomsky, theoretical linguist, characterizes ChatGPT as basically high-tech plagiarism that will lead to avoidance of learning \cite{b53}. Francois Chollet, the creator of the library Keras, has highlighted the lack of contextual reasoning in the use of the Large Language Models (LLMs), like ChatGPT \cite{b54}. Sam Altman, CEO of OpenAI, responded to these concerns by acknowledging the serious weaknesses and silly mistakes. He referred to ChatGPT as a very early glimpse of the changed world, attributed to Generative AI. He further acknowledged that there is still a lot to figure out \cite{b55}. The author in \cite{b1} concluded that ChatGPT can demonstrate critical thinking when solving problems, as the chatbot suggested enhancements to its own generated solutions. This was the basis for recommending the use of: multi-modal channels for exam questions, video recorded visual questions, AI output detection tools, and oral exams, to preserve the exam integrity. Another approach taken by some institutes is the complete ban of ChatGPT. For example, New York schools have cited the following reason for banning ChatGPT: it does not develop critical- thinking and problem-solving skills, which are essential for academic and lifelong success \cite{b56}. Similarly, a leading Dubai university \cite{b57} decided to reinstate pen and paper exams, after discovering that one third of students in a class were using LLMs to write their essays. Verena Rieser, professor at Heriot-Watt University, admits that despite being useful, ChatGPT can generate misinformation at a scale which is obviously very concerning, when it comes to education \cite{b58}. However, several other universities \cite{b59} have correctly recognized that the tool is here to stay and have moved away from discussions about outlawing it. The University of Bath will soon have a policy on ChatGPT and other Generative AI tools. The policy will address issues such as determining the extent to which productivity is enhanced, and ethical implications. While educational institutions are developing some policies and procedures, many staff at various universities are taking temporary measures such as setting in-person, invigilated summer exams, and/or asking students who choose to use ChatGPT to submit their ChatGPT prompts and answers as an appendix. Some tech giants and world leaders foresee the ChatGPT’s potential to revolutionize online interactions and education. Mark Zuckerberg, CEO of Facebook, is very optimistic about the applications of Generative AI in content generation, personalization, and virtual reality experiences \cite{b850}. Andy Jassy, Amazon CEO, during an interview has expressed his excitement to see “what is possible with Generative AI tools such as ChatGPT \cite{b60}” 

The education sector has not only successfully integrated calculators, Wikipedia, and essays-for-pay websites, but has also contributed to build most of these technology tools. If the past is any predictor of the future, then ChatGPT will have the same integration in the education sector. Many teachers now believe that ChatGPT could actually help make education better. 

Based on the public reaction to LLMs, we infer that the LLMs tools like ChatGPT are here to stay and in order to ensure the success of our students in the future workforce, we need to develop policies to integrate Generative AI tools in the higher education. The university policies may vary depending on the academic field and specific subject matter context; but should address the associated concerns about misinformation and bias, developing the students' cognitive capabilities and responsible AI. Currently, no universities have formal policies about LLMs usage; they have either blocked LLMs such as ChatGPT or remain passive. This article aims to i) assist academics take a clear position and ii) propose mechanisms for integrating LLMs into education.

\section{Education Crisis Prior to the Emergence of LLMs}\label{sec3}

Recent advances in LLMs have led many to question the role of education, with some suggesting that AI will replace human educators \cite{b195}. In this section, we recall the role of education in society. We also look at the different roles and learning objectives that distinct educational stages must achieve, as the integration of AI depends on the education stage and the learning outcomes associated with it.

The Delors report \cite{b5} delivered to UNESCO in 1996, is one of the most valuable sources that highlights the importance of education. This report was prepared by an international commission led by Jacques Delors, and its recommendations have been applied in 50 countries after being translated into 30 different languages \cite{b6}.

One of the key points of the Delors report is the importance of lifelong learning. Generative AI offers a unique opportunity to implement the Delors report recommendations, as it makes the learner less dependent on the formal education system and its educators. At the same time, however, it does not eliminate the need for the formal education system, whose role is not only to provide the basic knowledge, but also to define the best methods for integrating AI into education so that learners use these tools to build themselves up rather than over-relying on them.

Interestingly, throughout the whole Delors report, the international commission didn't provide any definition of what education is. Instead, the commission defined following the four pillars of education:

\begin{itemize}
    \item Learning to know: helps individuals to understand their environment and live in dignity. It enables individuals to develop their memory, reasoning, and critical thinking skills.
    \item Learning to do: allows individuals to apply what they have learned,
    \item Learning to live together: allows individuals to develop social skills that enable them to live and work with other individuals who might be different from them.
    \item Learning to be: teaches students to be creative, to form their own opinions, to criticize, and to have their own identity.
\end{itemize}


Furthermore, chapter 6 of the report discusses in details the differences between basic, secondary, and higher education. Basic education is presented in the report as '\textit{passport to life}'. It enables people to master essential learning tools such as literacy, oral expression, problem-solving,  building skills, values, and attitudes. Secondary education is presented in the report as a crucial point in the life of young students because, at this stage, they decide their future based on their tastes and what they acquired during their education. The role of higher education is to inculcate the culture of creativity in students. The report states that "[The role of] higher education [is]...to play its part in \textbf{creating}, \textbf{preserving and passing on knowledge} at the highest levels."

AI can be integrated into all stages of education. But its integration depends on the education stage, whether the objective is to enable students to master the basic knowledge, to help them decide on their future (secondary education), or to improve their creativity to face our constantly evolving world and the needs of the employment market (i.e., higher education). For example, the integration of AI in early childhood education must be carefully studied because the main objective at this stage is to focus on developing children's social skills. While the Delors report didn't explicitly specify the stages at which we expect the pillars of education to be applied, we obviously expect all the four pillars to be needed at the higher education stage. Indeed, the report states, "\textit{It is primarily the universities that unite all the traditional functions associated with advancing and transmitting knowledge: research, innovation, teaching and training, and continuing education.}"
Therefore, we believe that instructors in higher education should design their courses and exams to incorporate the four learning pillars. This will help develop students' critical thinking capacity and enable them to create and innovate for society, without forgetting the "Learning to be" pillar that helps students form their own identity and opinions.

However, the current higher education system suffers from many problems that hinder the achievement of the four pillars of learning. One of the main issues is the exclusive reliance on one single textbook for teaching. When professors are not the creators of the textbook, it becomes challenging for them to assist their students in building their own identities since they do not demonstrate this ability in their teaching (the "Learning to Be" pillar).  

In addition, while memorization is known to be one of the basic educational practices \cite{b5}, professors in higher education should avoid assessment methods that overly require memorization because, typically, students need to be prepared for the job market where companies provide employees resources (e.g., the Internet, books, chatbots, etc.) to solve problems at hand. The Delors report clarifies this issue by stating: "\textit{It would be dangerous to imagine that memory has become unnecessary because of the incredible capacity to store and circulate information now at our disposal. We must certainly be selective about what we learn 'by heart'}".

\begin{figure}
    \centering
    \includegraphics[scale=0.3]{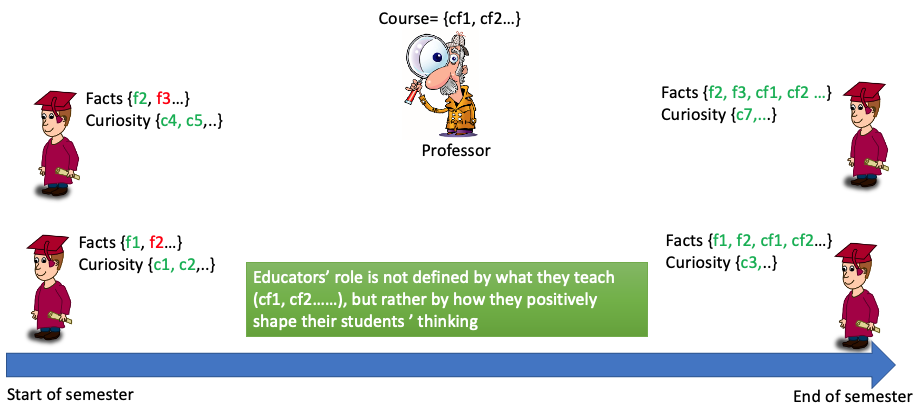}
    \caption{Personalised Learning}
    \label{fig0}
\end{figure}

Another problem that needs to be acknowledged is the focus on students' final outcomes and answers rather than their process of working through problems. The correctness of the final outcomes of students' work should not really be the main objective of educators because their role is to develop the intelligence of students. Improving the students' intelligence requires personalised learning. In 2015, the OECD prepared a report titled 'Schooling for Tomorrow – Personalising Education' \cite{b196}. In personalised learning, the educator should make an effort to understand the needs of every student, as they are typically different between students. Educators should also detect errors in the background knowledge of every student and work with each student to address their own curiosity and correct their previous knowledge while also delivering their course contents at the same time (see figure \ref{fig0}). The role of educators is not only defined by what they teach (i.e., the facts of their courses); but rather and more importantly by their ability to develop students' cognitive skills by addressing their students' curiosity and their misconceptions.

\begin{figure}
    \includegraphics[scale=0.4]{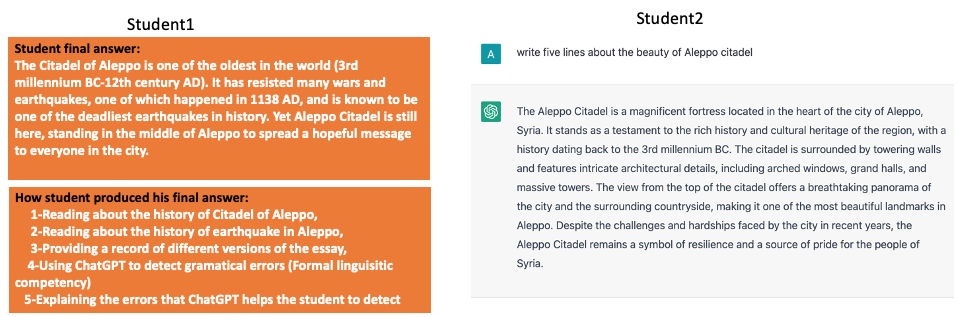}
    \caption{Example of English essay}
    \label{fig01}
\end{figure}

Consequently, all the evaluation rubrics should be redesigned to measure students' ability to think rather than focusing exclusively on the final answers that students provide during their course activities and exams. For example, figure \ref{fig01} illustrates an example where a student is asked to write five lines about the beauty of the Aleppo citadel. Student1 provides his teacher with his final answer (i.e., the essay) along with other pieces of evidence representing his efforts to produce the essay. In contrast, student2 provides only his final answer without giving evidence about how he/she produced the final solution. As it can be seen, student1 uses ChatGPT more thoughtfully and effectively than student2. Hence, the role of educators is to encourage the former chatbot usage while discouraging the later.

It is thus in the hands of every educator to avoid what the French philosopher Albert Camus once said: "\textit{The school prepares us for life in the world that does not exist}". The Delors report nicely summarises the attitude that every educator has to follow in the education systems: "\textit{Teachers' great strength lies in the example they set, of curiosity, open-mindedness, willingness to put their assumptions to the test and to acknowledge mistakes; most of all, they must transmit a love of learning.}" Reducing the roles of educators to rule enforcers and cheating deterrent does not help students to be critical thinkers. Instead, educators must entertain the culture of freedom and openness with students while maintaining in a seamless way the integrity of learning. 

\section{Adapting Course Design to AI's  Limitations and Strengths}\label{sec4}

A basic and simple definition of Artificial Intelligence(AI) is given by the creator of the AI concept, John McCarthy, in 1955 as "\textit{the science and engineering of making intelligent machines}" \cite{b34}.

Machine learning (ML) is the primary method used today to make machines intelligent. That's why, we often use the terms AI and ML interchangeably. The term "Generative AI" refers to the tools that implement ML techniques to generate new content. When the generated content is textual, we use the term "Large Language Models" (LLMs) to specifically refer to Generative AI for texts such as ChatGPT.

The recent media hype surrounding AI may give the impression that AI is the panacea of all kinds of problems in life, which is not true. It is better for educators to consider that the ideal use of AI is to complement human intelligence in order to understand the intelligence of the universe which humans don't (will never) completely understand. 

The strengths and limitations of AI can be boiled down to the study of the strengths and limitations of machine learning (ML), which includes all the Generative AI tools that educators are concerned about.

Therefore, educators should constantly monitor and understand the strengths\cite{b890} and limitations of ML. The strengths of ML allow educators to provide personalized education to their students. In this case, the role of educators is to ensure that students are correctly exploiting the strengths of AI for their self-growth, as opposed to over-relying on AI. On the other hand, educators should also constantly be aware of the limitations of ML during the design of their courses. 

We first discuss the limitations of AI and how these limitations can be exploited in course design. Then, we provide different examples to illustrate these limitations. To depict the strengths of AI for education, we present AI-based tools that have been integrated into education before the advent of ChatGPT. 

\subsection{Exploiting the Limitations of AI and ML for Course Design}
Machine learning (ML) is defined by IBM \cite{b7} as "\textit{Machine learning is a branch of artificial intelligence (AI) and computer science which focuses on the use of data and algorithms to \textbf{imitate} the way that humans learn, gradually improving its accuracy.}"

\begin{figure}
    \centering
    \includegraphics[scale=0.4]{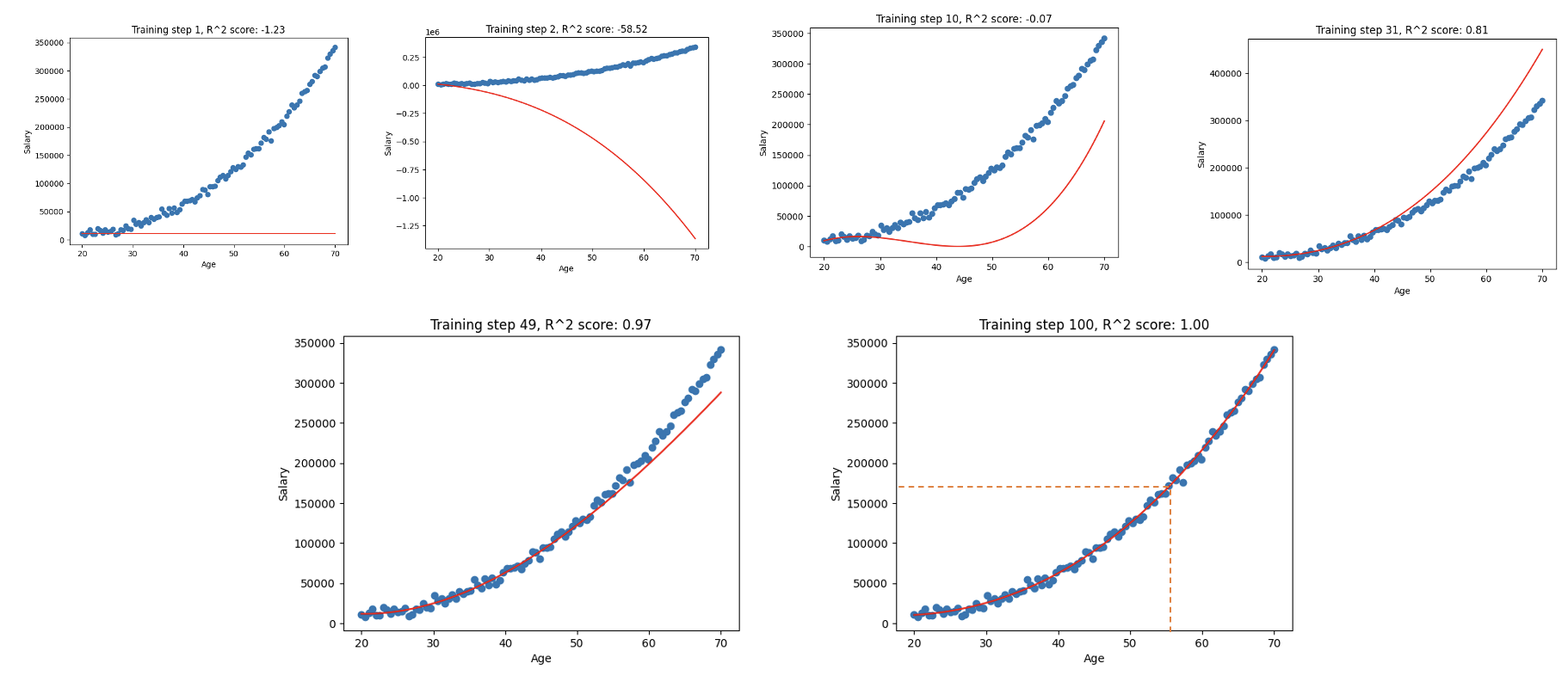}
    \caption{Example of training process}
    \label{fig1}
\end{figure}

Machines learn to imitate humans through a training process in which a dataset (which includes patterns of how humans solve problems) is used to enable the machine to identify patterns. Once the patterns are identified, the machine can imitate humans and predict future answers. For example, Figure \ref{fig1}  shows an excerpt of the training steps through which the machine attempts to identify the nonlinear pattern within the dataset. The dataset (depicted by the blue points) represents the relationship between salaries and the age of employees. The machine required 100 training steps to identify the pattern (illustrated by the red curve) in order to predict an employee's salary based on their age.

The first important limitation of ML is that it is probabilistic, thereby it is not suitable for solving problem that require deterministic answers, such as arithmetic operations or decrypting a message. In addition, patterns identified from a dataset may not be relevant because random phenomena contained in a dataset may never repeat in the future. The statistician Genevera Allen \cite{b880} stated: "\textit{But there are cases where discoveries aren't reproducible; the clusters discovered in one study are completely different than the clusters found in another. Why? Because most machine-learning techniques today always say, 'I found a group' Sometimes, it would be far more useful if they said, 'I think some of these are really grouped together, but I'm uncertain about these others.'}". Thus, a problem should be solved using ML only if it is not possible to solve it deterministically (e.g., recognising a handwritten digit) or if discoveries aren't reproducible (i.e. random).

The second limitation that educators and students must be aware of is that machines don't understand their output, because ML lacks common sense \cite{b8}. According to Gunning \cite{b9}, machine common sense has been always a critical-but-missing component of AI.  Teaching common sense to machines is very difficult because common sense is a set of unwritten rules that humans use to make sense of the world \cite{b8}. It is incredibly difficult for humans to think of and write these rules so that machines can learn them. There is nothing to suggest that machines understand their own generated output, which is generated using a taught pattern. Typically, we depend on humans to read and understand the output of machines. 

Thirdly, students and educators should recognise that ML datasets can be flawed and incomplete. The answers given by ML algorithms might not always be the correct representation of the real world problem at hand. There are several reasons for this, such as apophenia, bias and adversarial attacks. Apophenia refers to the mis-perception of patterns in random data \cite{b12}. For example, people may not like to travel on the 13th of the month or may come up with conspiracy theories. Even scientists may tend to define patterns for random phenomena, such as defining patterns of prime numbers. In terms of bias, any dataset is just a statistical sample that can never fully represent the complete view of the real world. Pasquinelli \cite{b13} defines three types of biases:
\begin{itemize}
    \item World bias: The world and society are already full of biased phenomena (e.g., gender, racial, and age biases), and they are independent of AI and any technology. However, with the increasing role of AI, there is a considerable risk of amplifying societal biases.
    \item Data bias: This bias can be introduced through the data capturing, formatting, and labeling, as well as the feature selection from the training dataset. Various factors can contribute to this bias, including the lack of diversity and knowledge among ML developers. 
    \item Algorithm bias: This bias can be introduced due to biased design in the optimization algorithms used to train the model. For instance, when hiring candidates for a job, a higher weight may be assigned to candidates from a specific university. In this case, the candidate selection process is not equitable.   
\end{itemize}

Malicious attackers are constantly evolving their strategies for corrupting datasets \cite{b14, b15, b16}. For example, Aryal et al. \cite{b14} discussed the use of a label-flipping approach to malign the training data of malware detectors. This technique allows attackers to bypass malware detectors to infect users' machines.  Thus, educators need to nurture critical thinking of their students, to enable them to counter such biases by criticising the partial and incomplete output of AI machines.

Another ML limitation that educators and students should be aware of is the responsibility of AI. Currently, the allocation of responsibilities for the outcomes of actions based on machine learning is not clear. In 2004, Andreas Matthias \cite{b17} discussed the issue of AI responsibility, referring to it as "The Responsibility Gap." Unfortunately, there is currently no efficient solution available to address this gap. Matthias emphasized that "\textit{Society must decide between ceasing to use this type of machine (which is not a realistic option) or facing an irresolvable responsibility gap that cannot be addressed using traditional concepts of assigning responsibility}." Recognizing this limitation is crucial for determining the usage policies of AI tools and for preparing students for their future careers in the market. 

ML suffers from another serious limitation, which is its inability to handle inputs from outside the data distribution used to train it. During training, datasets are typically divided into two sets: the training set and the validation set. The training set allows the machine to learn patterns, while the validation set is used to test whether the machine has learned the patterns accurately. However, both sets have the same data distribution. When confronted with new out of data distribution set scenarios, the machine either fails or demonstrates shallow performance. This problem is called the Out-of-Distribution (OOD) generalization problem in the literature \cite{b18}. Various examples can illustrate the OOD. For example, a machine cannot be used to recognize cats if it is trained on dog photos. Another example of OOD is data drifting  \cite{b19,b20,b21}. This issue arises when the underlying distribution of data has changed over time. For instance, house prices unexpectedly changed significantly due to the pandemic, which was not foreseen during ML training phase conducted prior to pandemic \cite{b20}. 

From an educational perspective, educators can always develop strategies to teach outside of the model's distribution. For example, the self-critique technique \cite{b21, b22} is one of the most advanced prompt engineering techniques that enables LLMs, such as ChatGPT, to assess and enhance their own output. Student can be encouraged to utilize this technique to answer exam questions or solve homework assignments. However, if an educator correctly designs the exam questions or the homework outside the distribution of the model, LLMs won't provide the correct answer to students. Table \ref{tab:1} presents a modified example from \cite{b23} that lies beyond the model's distribution. ChatGPT could not solve the problem because it was not trained to handle the summation of large numbers

\begin{table}
\centering
\small
\begin{tabular}{|c|p{10cm}|}
\hline
\multirow{2}{*}{Prompt} \\
& Valerie earns \$50005556677778888888 per month, 1/15 of what her brother earns. If their mother earns twice their combined salary, what's the total amount of money they all have together? \\
\hline
Answer & To find the total amount of money they all have together, we can first determine Valerie's brother's monthly earnings. Since Valerie earns 1/15 of what her brother earns, her brother must earn \$7,500,833,335,555,333 per month (15 * \$50005556677778888888).

Now, let's calculate their combined monthly earnings. Valerie and her brother's combined earnings amount to \$7,500,833,340,555,111,223 (\$50005556677778888888 + \$7,500,833,335,555,333).

According to the information provided, their mother earns twice their combined salary, which is \$15,001,666,681,110,222,446 (2 * \$7,500,833,340,555,111,223).

To determine the total amount of money they all have together, we add up their individual earnings: Valerie's earnings (\$50005556677778888888), her brother's earnings (\$7,500,833,335,555,333), and their mother's earnings (\$15,001,666,681,110,222,446). The total amount of money they all have together is \textbf{\$15,001,667,132,667,667,667} (\$50005556677778888888 + \$7,500,833,335,555,333 + \$15,001,666,681,110,222,446).

\textbf{Critique and Improvement}:

The initial answer accurately calculates Valerie's brother's earnings and the combined earnings of Valerie and her brother. However, there seems to be an error in calculating their mother's earnings and the subsequent total amount of money they all have together.

To improve the answer, let's correct the calculations:

Valerie's earnings: \$50005556677778888888
Her brother's earnings: \$7,500,833,335,555,333

To find their combined earnings, we add Valerie's earnings and her brother's earnings: \$50005556677778888888 + \$7,500,833,335,555,333 = \$8,000,839,892,133,222,221.

According to the information provided, their mother earns twice their combined salary. Therefore, their mother's earnings would be \$16,001,679,784,266,444,442 (2 * \$8,000,839,892,133,222,221).

To calculate the total amount of money they all have together, we add up their individual earnings: Valerie's earnings (\$50005556677778888888), her brother's earnings (\$7,500,833,335,555,333), and their mother's earnings (\$16,001,679,784,266,444,442).

The improved calculation gives us a total amount of \$24,001,713,818,620,777,663 (\$50005556677778888888 + \$7,500,833,335,555,333 + \$16,001,679,784,266,444,442).

Therefore, the total amount of money they all have together is \textbf{\$24,001,713,818,620,777,66}3. \\
\hline
\end{tabular}
\caption{Example modified from \cite{b23} to be out of distribution}
\label{tab:1}
\end{table}

Finally, the last point that needs to be addressed by educators is the current debate about whether AI machines are capable of thinking or not. This debate dates back to 1965 when the author of ELIZA chatbot wrote a book \cite{b24} illustrating how patients thought of his chatbot as a human psychotherapist despite the fact that it only used basic text processing techniques . It is important for educators to follow the arguments of both camps because the role of education is to develop human intelligence. Educators should ensure that students are always able to develop deep logical reasoning, that surpasses the machine's superficial pattern based capabilities in their respective learning areas.

We strongly believe that AI machines cannot think because of their dependence on humans to perceive and analyze world problems. Machines can only truly think if they are capable of perceiving the world autonomously and defining patterns to problem-solving without the help of humans. Yann LeCun, one of the grandfathers of AI, wrote a position paper \cite{b25} in which he illustrates the modules that need to be implemented by machines to replace humans, and one of the modules is the world's perception.  

Educators have to know that there is currently no official method for proving whether a machine is intelligent or not. An accepted method is to select a reference dataset and test all models against this dataset. However, anyone can define a test outside of the reference dataset, thereby stumping the machine's problem solving capabilities. Additionally, there is no scientific method that explains the acceptable error rate resulting from training processes; it is purely the decision of the machine learning trainer \cite{b13}. Turing defined another method of evaluating machine intelligence through observation \cite{b10}. Machines are considered intelligent if a human observer is unable to differentiate its output from that of another human. However, there is a debate in the literature regarding the relevance of the Turing test to evaluate the intelligence of machines \cite{b26}.  

Consequently, educators should design their courses and exams based on the limitations and debates that are surrounding the AI and ML technology. In the higher education, most of educators are also researchers. The research knowledge of educators must be used to design courses in the AI era because this knowledge allows educators to go beyond the limitations of AI machines enabling their students to use AI to complement the human intelligence in understanding the universal intelligence.

\begin{figure}
    \centering
    \includegraphics[scale=0.4]{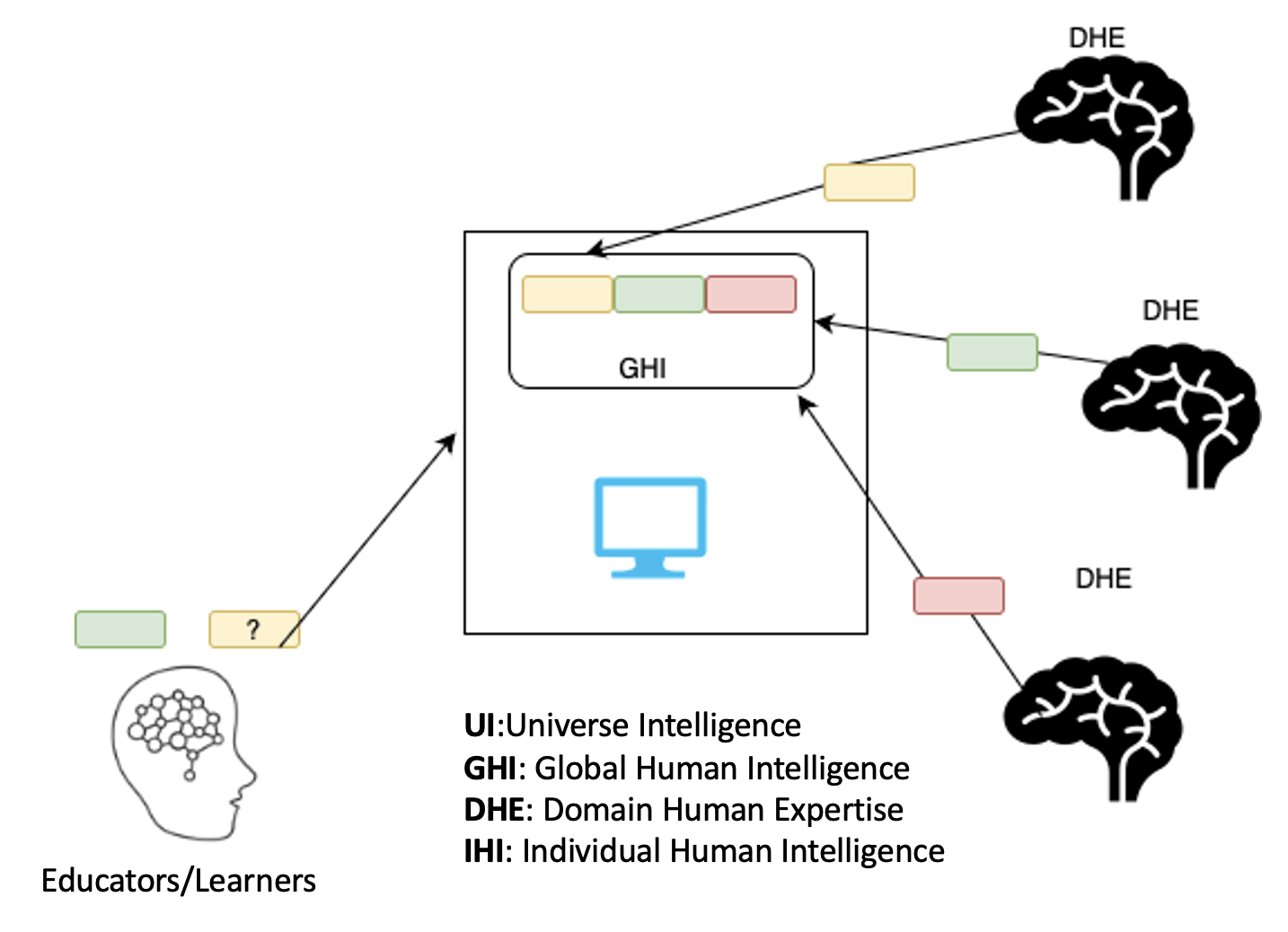}
    \caption{AI machines store the Global Human Intelligence that is very far from universe Intelligence}
    \label{fig:04}
\end{figure}

Ultimately, educators should consider AI machines as tools capable of performing tasks similar to those accomplished by human experts, such as drawing, reasoning, and writing. When educators possess expertise in a specific domain (e.g., the green domain shown in Figure \ref{fig:04}), they will recognize the limitations of AI machines in that particular area. However, it is important to note that educators and students cannot be experts in all domains and areas (e.g., the yellow and red domains in Figure \ref{fig:04}). Therefore, AI can assist educators and students in developing Individual Human Intelligence (IHI) in areas where their understanding is incomplete. Global Human Intelligence (GHI) is ultimately the sum of all Domain Human Expertise (DHE) that can help develop the IHI of every individual. Thus, educators in the higher education have the responsibility to guide students in their area of expertise (i.e., the green domain), where the use of AI is safe because the educator's knowledge surpasses that of the AI machine.

It is worth noting that the DHE of the universe, whether considered indeterministic like Heisenberg \cite{b194} or deterministic like Einstein \cite{b193}, is never complete or absolutely correct. Indeterminists believe that scientific theories can never be proven to be absolutely true but can only be falsified. Thus, for indeterminists, the role of scientific discovery is to refine and improve theories. On the other hand, determinists believe that universal events can be predictable if we know all the past events and the laws of nature \cite{b192}. However, they acknowledge the existence of hidden variables that are challenging to discover. Therefore, the role of scientific discovery is to uncover the universe's hidden states. Both views of the universe highlight humans' limitations in understanding it. As educators, it is our role to define pedagogical approaches that allow us to enhance our understanding of the universe. By incorporating AI in their courses, educators can help humanity improve its understanding of the universe and accelerate science discovery. 

Furthermore, it is very important for educators to establish an identity in their respective teaching areas because their role is to teach their students the "Learning to be" pillar. Based on the relativism theory in Philosophy, all the truth and falsity, good and bad are not absolute concepts but rather based on the culture and context that these values are defined \cite{b27}. Consequently, educators can adapt all the knowledge of textbooks they teach to the context and culture of students. By doing so, educators can help students build the capacity to critically analyse the knowledge of their textbooks from their own cultures and perspectives.

The role of educators is to perpetually observe the limitations and strengths of AI in order to design their courses in a way that enable their students to use any technology that augment their human intelligence. We know that machines are not thinking, so our role is to make our students think. We know that AI machines hallucinate and have no common sense, then it is our role as educators to develop in our students the capacity to check facts and demonstrate understanding of machines' output. 

\subsection{Exploiting the Strengths of AI Chatbots for Education}
Even before November 2022, the launch date of ChatGPT, chatbot integration into education has been an active research domain in the field of AI in education; with the scientific community contributing to the three broad directions: 
\begin{itemize}
    \item determining the chatbot integration impact on education;
    \item the techniques to build chatbots; and
    \item the adaptation of chatbots according to the students’ psychology.
\end{itemize}
Educational chatbots could be used as powerful classroom aids in various capacities such as evaluator, tutor, responder, moderator, and peer learner \cite{b85}. ELIZA \cite{b61} ,despite its primitive design,  has been used to understand the educational impact of chatbots \cite{b62}, only to conclude that such trivial design chatbots had no significant impact on the students’ critical thinking. This hypothesis was further supported by \cite{b63} using a chatbot that guides the students in essay writing with the stipulation that it is always the chatbot – as opposed to the student – that would initiate the discussion. More than 30 years after ELIZA, ALICE was developed using AIML and the chatbot response was based on the matching of the pre-stored classification patterns \cite{b64}. Abbasi \cite{b65} uses this same strategy of picking a response from a pool of pre-stored classification patterns to develop a knowledge base of 5000 questions and categorizes them into 7 categories: what, why, who, how, perform operations, application, advantage/disadvantage. Despite having a simple retrieval operation, the students had a better memory retention when compared with Google using the same pre-defined  questions, for the course of Object-Oriented Programming. The chatbot prototype functionality was enhanced by adding text to speech conversion system and the performance was measured using 2 criteria: capacity measurement on the students’ cognition capabilities and the quality of the user interface \cite{b66}. The same technique of using pre-programmed questions with an extendable knowledge base was used by Bii \cite{b67} to understand the teachers’ attitude towards chatbot usage in education. The derived chatbot from open source resources \cite{b68} was named KNOWIE and was also used to assess students' attitude \cite{b69} and seek improvement suggestions for the chatbots if they are to deploy in their institution. In a similar work, Na-Young \cite{b70} determines the positive effects of chatbot usage while improving English grammar skills.
Instead of picking a response from a pool of pre-made answers, the authors in \cite{b71} explore the self-paced chatbot mediated learning (CML), focused on the individual need of the student. In March 2022 the self-based CML was taken to another level when Quizlet added a ChatGPT built feature Q-Chat \cite{b72} that adjusts the complexity of the questions based on students’ needs and preferences. So, humanities and computer science students will have different experiences with the CML, using such advanced chatbots. Perez-Marin \cite{b73} refers to these AI chatbots as Pedagogic Conversation Agents (PCA), which create an interface that acts as instructor, companion, or learner by teaching. Based on the assumed role, ten design criteria were suggested, such as usage within cell phones or computers, generic knowledge or subject specific knowledge and then suggested concept map creation based on the design criteria. Hobert \cite{b74} evaluates the PCA benchmarks using a 4-step process to conclude that PCAs using the interfaces of instant messenger apps  such as WhatsApp, Facebook are more popular among students. He then uses Spearman’s rank correlation coefficient \cite{b75} to conclude that small talk conversations are supportive in long term adaptation of the chatbot based learning system. Hobert further analyzes \cite{b76} the research work to come up with a 3-point research agenda for chatbot PCA usage in education – generalized design knowledge, comprehensive evaluation benchmarks, and process models that cover both: design steps and evaluation methods.  Using the same motivation as \cite{b76}, that the cell phones provide a personalized environment for the students, Carayannopoulos \cite{b77} implements a motivational companion chatbot called BU111 on the Kik messaging platform. Besides sending the motivational reminders regarding the weekly assignments, BU111 was also used by Instructors to send the reminders and encouragement notes to the students. The weekly assignment reminders disappeared gradually as the term progressed, as the students were reported to be more autonomous by the end of the term. The selection of a chatbot platform (e.g. Chatfuel, Botsify) is an important step of the chatbot development, as defined by Davies \cite{b78}. He then integrates a chatbot that uses Amazon Web Services (AWS) with an e-Learning system of English language courses, to help students in their home exercises based on their psychological preference; such as bi-directional communication or lecture oriented based conversation. Such student learning preferences are stored in DynamoDB which is accessible by the corresponding learner using Amazon Cognito secure authentication. Davies concluded that chatbot cannot replace real instructors, when it comes to meeting personalised learning needs of students. In another effort, IBM Watson Assistant was used to implement three text-based chatbots, in a flipped classroom setting. All three were named 7345 bot to create an illusion of a single chatbot for the students \cite{b79}. Similar to the conclusions of Davies, it was  concluded that students treat chatbots as a mere virtual character; probably because of the 7345 bot’s inability to answer any open-ended questions and therefore no solution for unstructured problems. Matsuura \cite{b80} created a chatbot called NAO on the platform Api.ai and connected it to an online learning system based on Topic Maps, thereby developing a terminology visualization that supplemented instructor’s verbal explanation by providing visual explanation of the terms. However, miss-recognition and inappropriate display happened frequently, causing the loss of concentration for the students. Kerly \cite{b81} employs Wizard of Oz method and shares several lessons learnt regarding chatbot design requirements: chatbot should have links to external databases, need for small talk, preventing the chatbot loss by ensuring that the student does not close the chatbot window. To this end, we can see that there has been a lot of effort in measuring the chatbot integration suitability in the education sector, and both: the students and instructors have favorable views of deploying chatbots in the education sector, to achieve one goal or another.
Chang \cite{b82} categorizes the chatbot designs in three categories: knowledge base that uses established rules; retrieval based where answers are retrieved based on the similarity of the pre-stored question and question being asked; and Recurrent Neural Network based, where two Long Short Term Memory Networks act as encoder and decoder to process the questions and answers in the memory, respectively. He then constructs a dialogue system based on first and third models and tests the results of the chatbot design on college students to conclude that the students have positive views of instant messaging exchange with chatbots based on the combination of multiple design approaches. As pointed out by Almahri \cite{b83}, the perceived likeliness of adopting chatbots in a university learning environment is dependent on 4 key constructs: performance expectancy, effort expectancy, social influence, and facilitating conditions. The behavioral intentions determine the actual use of technology, and this theory is referred to as Unified Theory of Acceptance and Use of Technology (UTAUT). The behavioural intentions differ from one cohort of students to the other, e.g. computer science students' adaptability curve to accept chatbots would significantly differ from fine arts students. A case study that supports this theory was carried out by Fryer \cite{b86} where the behaviour intentions were generalised. e.g. it was assumed that the students feel more relaxed when talking to a machine than a person. While this assumption could be true for foreign language learning, the students pursuing a degree in computer science might feel differently. 
Even in the ChatGPT dawn era, the knowledge that the instructors expect the students to acquire has stayed the same. However, it will be convenient for the students to find the correct information if they are trained to ask the right questions, as this will facilitate them to comprehend and analyse the answers. The role of instructors -besides teaching - is also defined by helping students to enhance their reasoning capabilities using the available technology tools of our era, such as ChatGPT. Therefore, it is imperative that the instructors develop new pedagogical approaches that incorporate AI chatbots such as ChatGPT in their teaching approach. In the absence of such a pedagogical approach, AI tools such as ChatGPT will do more harm than good, by encouraging students to cheat on the answers. Given that schools have survived calculators, Google, Wikipedia, essays-for-pay websites, and more, this work analyzes the behavioral patterns of the students while using ChatGPT and based on the analyses, proposes a pedagogical approach that entails ChatGPT usage in education.

\section{What Educators Need to Know About Large Language Models (LLMs)}\label{sec5}

LLMs are the tools that have the most impact on the education system. Every educator must demonstrate a high level of understanding of the internal workings of LLMs so that they can properly evaluate, critique students' work, and understand that LLMs (and ML in general) are purely statistical-based generation machines, not capable of thinking. 

\begin{figure}
    \centering
    \includegraphics[scale=0.5]{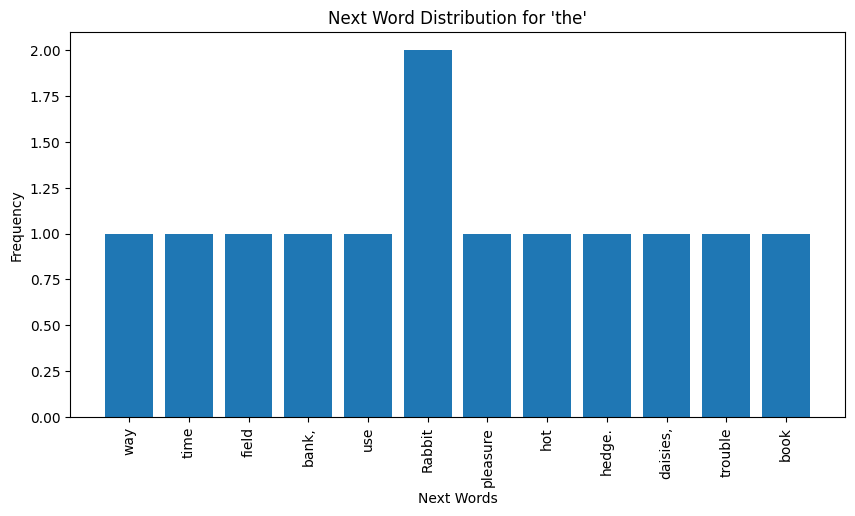}
    \caption{List of possible words can be used after "the" according to our dataset}
    \label{fig:041}
\end{figure}

From a mathematical point of view, LLMs are predictors of the next word (or token) given a context (provided by the user) that is composed of a set of words (or tokens). For example, Table \ref{tab:2} shows generated text using a Markov chain which enables generating the next word based on the previous word. The generated text depends on the text provided as input to the Markov chain, which allows defining the distribution of each word. For instance, figure \ref{fig:041} shows the list of possible words that come after the word "the". In the generated text, the word "trouble" is selected. Once the next word is chosen ("trouble"), the list of possible words after the selected word ("of") is considered until generating the complete sentence.

\begin{table}
\centering
\small
\begin{tabular}{|c|p{10cm}|}
\hline
\multirow{2}{*}{Dataset} \\
& Alice was beginning to get very tired of sitting by her sister on the bank, and of having nothing to do: once or twice she had peeped into the book her sister was reading, but it had no pictures or conversations in it, 'and what is the use of a book,' thought Alice 'without pictures or conversations?'

So she was considering in her own mind (as well as she could, for the hot day made her feel very sleepy and stupid), whether the pleasure of making a daisy-chain would be worth the trouble of getting up and picking the daisies, when suddenly a White Rabbit with pink eyes ran close by her.

There was nothing so very remarkable in that; nor did Alice think it so very much out of the way to hear the Rabbit say to itself, 'Oh dear! Oh dear! I shall be late!' (when she thought it over afterwards, it occurred to her that she ought to have wondered at this, but at the time it all seemed quite natural); but when the Rabbit actually took a watch out of its waistcoat-pocket, and looked at it, and then hurried on, Alice started to her feet, for it flashed across her mind that she had never before seen a rabbit with either a waistcoat-pocket, or a watch to take out of it, and burning with curiosity, she ran across the field after it, and fortunately was just in time to see it pop into a rabbit-hole under the hedge. \\
\hline
Generated text &  the trouble of the field after it, and looked at it, and burning with pink eyes ran close by her own mind that she thought it occurred to get very tired\\
\hline
\end{tabular}
\caption{Text generated using Markov chain to produce 30 words sentences that start with the word "the"}
\label{tab:2}
\end{table}

The main problem of the Markov Chain-based method is that it doesn't capture the context comprehensively and can generate a text that doesn't make any sense (e.g., "pink eyes" in our generated text). The n-gram method \cite{b901} allows us to generate a better text by considering \textit{n} previous words rather than only the previous word, as in the Markov Chain. For instance, in our previous example, if we adopt the 3-gram method, the probability of words is calculated based on the previous 2 words. For example, the probability of the word "the" can be calculated as follows:  $$P("the" | "be\;worth")$$

N-gram methods become impractical when \textit{n} (the number of words to be considered) becomes large. It becomes almost impossible to find a set of words when \textit{n} is large. In practice, 5-gram is commonly used because any larger value of \textit{n} doesn't improve prediction quality significantly.

In 2003, an important breakthrough happened in the field of text generation through the work of Bengio et al. \cite{b28}. Bengio et al. proposed using a neural probabilistic language model that outperformed n-gram models by utilizing neural networks. The main idea is that instead of calculating the probability of the next words based on their occurrences in the dataset, semantic and syntactic dependencies of the word to predict can be calculated based on trained neural networks. The term \textit{word embedding} is defined later to refer to this process. The neural probabilistic language model captures a larger context than n-gram and provides more meaningful generated text. 

In 2017, another important breakthrough happened through the work of Vaswani et al. \cite{b29} who proposed a new neural architecture that outperform Recurrent Neural networks (RNN) used by Bengio et al. \cite{b28}. Vaswani et al. \cite{b29} proposed a simpler and more efficient neural architecture called Transformer that uses a mechanism (called attention mechanism) that allows to focus on some parts of the contextual words used to predict the next word. This breakthrough allows the creation of what is called today Large Language Models (LLMs) such as GPT, GPT-2, GPT-3 and GPT-4. These models are trained on large data using self-supervised learning. The term  Foundation models is used to refer to this models \cite{b30}. These models can be used for various downstream tasks (such as question/answering, sentiment analysis and instruction following) \cite{b30} using a specific process called fine tuning. For example, Vicuna results from fine tuning LLama model from Meta using dataset obtained from sharedgpt.com.

It is important to understand different facts about the foundation LLMs models \cite{b31}: 

\begin{itemize}
    \item LLMs are getting more capable with increasing investment, not with targeted innovation. According to the author \cite{b31}, the main difference between GPT, GPT-2 and GPT-3 is not the architecture but the investment made in the training in term of data and computing power.
    \item  It is not possible to predict when specific skills or capacities will be exhibited by LLMs. In many cases, a model can fail prior to suddenly becoming capable when trained at better scale.
    \item There are no reliable techniques to control the behavior of LLMs. Different methods such as fine tuning or reinforcement learning can allow to improve the behaviour, but there is no guarantee that the model will behave appropriately in every situation.
    \item LLMs are using neural networks with billions of connections between its artificial neurons. It is almost impossible to understand precisely the behavior of such complex systems \cite{b900}.
\end{itemize}
Thus, educators must be aware that there are very few opportunities for students to customize the output of LLMs. Students would not be able to fully control the generated text. Within that framework, we discuss the pedagogical approach that can be followed to integrate LLMs in class and exams in higher education, in the following section.

\section{Courses Delivery and Exams for the Higher Education}\label{sec6}
If we want to integrate AI into higher education courses and exams, we need to take into account the four types of learning presented in the Delors report, discussed in Section \ref{sec3}. In addition, we have to consider the strengths and limitations of AI tools we detailed in Section \ref{sec4} to ensure that answers cannot be obtained by simple click of buttons.


It is very important to train students for the correct usage of LLMs before exams. During every class session, professors can encourage their students to use LLMs to facilitate class discussions and assignments. This practice would allow students to become familiar with LLMs and recognize their benefits and drawbacks in handling class concepts.

However, the traditional teaching methods such as lecturing might not be the best teaching approach that educators can follow to help their students master the use of LLMs. The Socratic teaching approach \cite{b32} appears to be a very engaging method that educators in higher education can apply with their students. The core idea of Socratic teaching approach is summarised as follows \cite{b32,b33} : "\textit{Accordingly he asked questions, letting the other man do most of the talking, but keeping the course of the conversation under his control, and so would expose the inadequacy of the proposed definition of courage. The other would fall back on a fresh or modified definition, and so the process would go on, with or without final success}". 

The Socratic approach enhances students' critical thinking skills. It requires educators to ask open questions and guide students in analyzing class questions. The use of LLMs enable students to interact individually with chatbots for analyzing the questions posed by the educators. Educators must design class questions in a way that prevents simple button clicks from providing the answers. Instead, answers should be obtained through a series of follow-up questions asked by students.

Similarly, professors have to ensure that the answers of exams are not obtained by simple click buttons, but instead are the results of deep reflections made by students during their exams. For example, LLMs can be used by professors to generate exam questions and by students to prepare themselves for their exams. However, this type of LLM's usage that we refer to as symmetric usage is not recommended because most of the questions will be answered perfectly by the LLMs. For instance, when we asked ChatGPT to generate 10 Multiple choices questions for exams in three different areas ( OS, Network security and Math), ChatGPT (using different user's account) was able to obtain a score that ranges from 80\% to 100\%.

If professors modify the questions generated by LLMs, it will be more challenging for students to obtain high grades. Thus, the most suitable usage of LLMs is the asymmetric usage of LLMs. This can happen in two different ways: (1) professors generate exams using LLMs and prevent students from using LLMs during their exams or (2) professors generate their exams using their own knowledge and background and allow students to use LLMs during their exams. 

We believe that the first method is more suitable for basic and secondary education in which students have to learn and master different basic concepts that are necessary for their higher level studies in the university. As Delors stated in his report \cite{b5} "\textit{although people need to take every opportunity for learning and self-improvement, they will not be able to make good use of all these potential resources unless they have received a sound basic education.}"

The second method is more suitable for higher education in which students have to be prepared to be more creative to face the unknowns of the market or the research. Clearly, the second method is more challenging for professors to write their exams, but this method allows students to enhance their critical thinking.

Any truth in life can always be obtained by asking a set of ordered and linked questions. Every answer to every question must be interpreted correctly to ask the next question until reaching the truth or the final outcome. Consequently, any class discussions, assignment or exam questions must be built on the idea that the final answer (essay, answers, drawing, etc.) must be obtained after a set of questions asked by the students.

\begin{figure}
    \centering
    \includegraphics[scale=0.5]{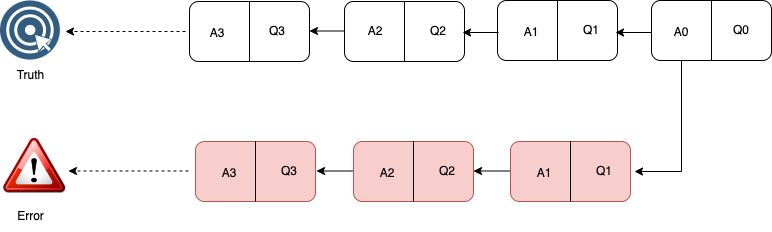}
    \caption{Questions Graph}
    \label{fig:truthgraph}
\end{figure}

In addition, educators can easily know whether the work returned by their students was the results of their own efforts or obtained from the LLMs using simple click buttons. Educators have to ask students to provide not the final answer but also a complete track log of all the actions that students applied to obtain their final answers. 

Thus the questions, assessments, projects and analytical queries must be carefully designed to be open-ended and based on class discussions. This way, the educator's questions can only be handled by those students who attend the class and pay attention to the class conversation. It is very difficult for LLMs to answer any question that is based on the class discussions, especially when educators design their questions based on the limitations of ML, explored in Section \ref{sec4}. Even when students manage to fine-tune LLMs using the class discussions, educators can always propose open questions based on the class discussion.

\section{Case Studies}
\label{sec7}

We provide in this section different exercises in four different areas: Cybersecurity, Programming, English language teaching and Art. Our objective is to explain whether the exercise should be validated or not from educational point of view. The main ideas that we want to reflect in these exercises:
\begin{itemize}
    \item Exercises must be open. Students must ask multiple follow-up questions before giving their final answer,
    \item Exercises must be based on the opinion of the educator but not on the textbook,
    \item Evaluation of exercises must be based on the effort of students to handle the exercise, not only the final answer.
\end{itemize}

Tables \ref{tab:5} and \ref{tab:6} demonstrate exercises that are not valid from an educational standpoint, whereas the Tables \ref{tab:7}, \ref{tab:9}, \ref{tab:10}, \ref{tab:12} and \ref{tab:13} show exercises that are valid from an educational one because they demand significant effort from the students before they can be solved.

\begin{table}[H]
\centering
\small
\begin{tabular}{ | m{8em} | m{10cm}| }
\hline
\textbf{Exercise} & Write a paragraph of 5 lines about the beauty of Aleppo citadel  \\
\hline
\textbf{ChatGPT answer} & 
The Aleppo Citadel stands as a magnificent testament to the timeless beauty of architectural mastery. Its towering stone walls command attention and inspire awe. As the sun casts its golden glow upon the fortress, the intricate details of its ancient design come to life, captivating all who behold it. From its imposing entrance to the delicate carvings adorning its ramparts, every corner of the citadel exudes a sense of history and grandeur. It serves as a living link to the past, reminding us of the enduring allure of Aleppo's rich cultural heritage.
\\
\hline
\textbf{Validation from education
 point of view}&  The essay is very well written from English point of view,  but the answer is obtained from a simple click. There is no learning effort from the student then the exercise is invalid from education point of view.\\
\hline
\end{tabular}
\caption{\centering Example English Essay}
\label{tab:5}
\end{table}

\begin{table}[H]
\centering
\small
\begin{tabular}{ | m{8em} | m{10cm}| }
\hline
\textbf{Exercise} & Write python script that encrypts a text using Caesar cipher  \\
\hline
\textbf{ChatGPT answer} & 
\begin{verbatim}
def caesar_cipher_encrypt(text, shift):
    encrypted_text = ""
    for char in text:
        if char.isalpha():
            if char.isupper():
                ascii_start = ord('A')
            else:
                ascii_start = ord('a')              
            encrypted_char = chr((ord(char) - 
                            ascii_start + shift) 
                            % 26 + ascii_start)
            encrypted_text += encrypted_char
        else:
            encrypted_text += char
    return encrypted_text
# Test the function
plaintext = input("Enter the text to encrypt: ")
shift_amount = int(input("Enter the shift amount: "))
encrypted_text = caesar_cipher_encrypt(plaintext, shift_amount)
print("Encrypted text:", encrypted_text)
\end{verbatim}
\\
\hline
\textbf{
educational validity}&  The obtained code is correct,  but it is obtained from a simple click. There is no learning effort from the student then the exercise is invalid from education point of view.\\
\hline
\end{tabular}
\caption{\centering Programming example}
\label{tab:6}
\end{table}

Examples given in Tables \ref{tab:5} and \ref{tab:6} illustrate the need for the education system to shift its focus from evaluating the final answers of students to the process followed by the students to reach their final answers. We now provide different examples that can be considered valid from education point of view. For each exercise, we explain the reason that makes it valid from educational point of view. 

\begin{table}[H]
\centering
\small
\begin{tabular}{ | m{5em} | m{11cm}| }
\hline
\textbf{Exercise} & 
\includegraphics[scale=0.5]{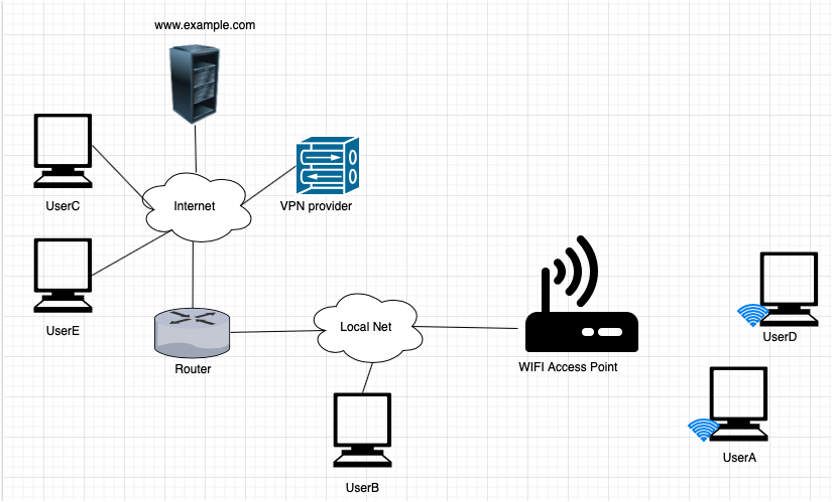}

UserE can see the traffic between the router and the VPN provider UserC can see the traffic between VPN provider and www.example.com UserB can see the traffic between the Access point and the router Who can understand or modify the messages of UserD o the server www.example.com, in the following scenarios:
Scenario1: UserD use the protocol HTTP UserD uses WPA3 to protect its WIFI communications
Scenario2: UserD use the protocol HTTPS UserD uses WPA3 to protect its WIFI communications
Scenario3: UserD uses the protocol HTTP UserD doesn’t use any protocol to protect its WIFI communications
UserD creates a VPN connection with the VPN provider
 \\
\hline
\textbf{ChatGPT answer} & 

Scenario 1: UserD uses HTTP protocol and WPA3 for Wi-Fi communications.

UserB (can see traffic between access point and router) cannot understand or modify UserD's messages.
UserC (can see traffic between VPN provider and www.example.com) cannot understand or modify UserD's messages.
Only UserD and the server www.example.com can understand or modify the messages.

Scenario 2: UserD uses HTTPS protocol and WPA3 for Wi-Fi communications.

UserB cannot understand or modify UserD's messages.
UserC cannot understand or modify UserD's messages.
Only UserD and the server www.example.com can understand or modify the messages.

Scenario 3: UserD uses HTTP protocol and no protection for Wi-Fi communications.

UserB can understand and potentially modify UserD's messages.
UserC can understand and potentially modify UserD's messages.
UserD's messages to the server www.example.com can be understood or modified by UserB and UserC.
Using encryption protocols like HTTPS and WPA3 is crucial to ensure the confidentiality and integrity of data transmitted over networks.

\\
\hline
\textbf{educational validity}&  The answer of ChatGPT is not correct when the student copies and pastes the question without extending it based on the associated image. This question is valid from education point of view because student needs to make a learning effort by reformulating correctly the exercise's question based on the associated image.\\
\hline
\end{tabular}
\caption{\centering Better example in Cybersecurity}
\label{tab:7}
\end{table}

\begin{table}[H]
\centering
\small
\begin{tabular}{ | m{5em} | m{10cm}| }
\hline
\textbf{Exercise} & Write a paragraph of 5 lines about the beauty of Aleppo citadel  \\
\hline
\textbf{ChatGPT answer} & 
The Aleppo Citadel stands as a magnificent testament to the timeless beauty of architectural mastery. Its towering stone walls command attention and inspire awe. As the sun casts its golden glow upon the fortress, the intricate details of its ancient design come to life, captivating all who behold it. From its imposing entrance to the delicate carvings adorning its ramparts, every corner of the citadel exudes a sense of history and grandeur. It serves as a living link to the past, reminding us of the enduring allure of Aleppo's rich cultural heritage.
\\
\hline
\textbf{Human answer} & 
The Citadel of Aleppo is one of the oldest in the world (3rd millennium BC-12th century AD). It has resisted many wars and earthquakes, one of which happened in 1138 AD, and is known to be one of the deadliest earthquakes in history. Yet Aleppo Citadel is still here, standing in the middle of Aleppo to spread a hopeful message to everyone in the city.
\\
\hline
\textbf{Efforts to produce human answer} & 
Students with his final answer shows evidence of :

1-Reading about the history of Citadel of Aleppo,

2-Reading about the history of earthquake in Aleppo,

3-Providing a record of different versions of the essay,

4-Using ChatGPT to detect grammatical errors (Formal linguistic competency)

5-Explaining the errors that ChatGPT helps the student to detect

\\
\hline
\textbf{Educational validation}&  The human student answer is not as perfect as ChatGPT answer. However, any student who gives an answer must be evaluated better than any student who is not providing evidence or can not explain his final answer.\\
\hline
\end{tabular}
\caption{\centering Better example of English Essay}
\label{tab:9}
\end{table}

\begin{table}[H]
\centering
\small
\begin{tabular}{ | m{5em} | m{10cm}| }
\hline
\textbf{Exercise} & The following question is encrypted by Caesar cipher. The used key corresponds to the number of students who attended the last session on campus. You need to test all possible keys if you don't know the number of students who attended the last class session on campus. You will obtain half of the grade if you decrypt the question. The other half of the grade will be given if you answer the question correctly.

The encrypted question

Hasodlq zkb fdofxodwlqj WFS vwdwh eb d vwdwhixo iluhzdoo lv pruh uholdeoh wkdq XGS vwdwh  \\
\hline
\textbf{ChatGPT answer} & 
To decrypt the Caesar cipher, we need to try all possible keys, which in this case correspond to the number of students who attended the last session on campus. We'll start by trying all numbers from 0 to 25 as potential keys.

The encrypted question is: "Hasodlq zkb fdofxodwlqj WFS vwdwh eb d vwdwhixo iluhzdoo lv pruh uholdeoh wkdq XGS vwdwh."

Using a Caesar cipher decryption tool or by manually shifting the letters, we can decrypt the question with each possible key. Let's go through the process for a few examples:

Key = 0: "Hasodlq zkb fdofxodwlqj WFS vwdwh eb d vwdwhixo iluhzdoo lv pruh uholdeoh wkdq XGS vwdwh." (No shift, original text)

Key = 1: "Gzrnckp yja ecnewncvki IER uvcvg da c uvcvgjwnh hktygcnn ku oqtg tgnkcdng vjcp WFR uvcvg."

Key = 2: "Fyqmbjo xiz dbmdvmbujh HDQ tubuf cz b tubufivmg gjsxfbmm jt npsf sfmjbcmf uibo VEQ tubuf."

Key = 3: "Expialm why calmulatig GCP state by a statehulfl firesalli is more reliazle than UDP state."

Continuing this process for all possible keys, we will find the decryption that makes sense in English. Once we decrypt the question, we can proceed to answer it to receive the remaining half of the grade.
\\
\hline
\textbf{Educational validation}&  The question exploits the fact that ChatGPT is probabilistic and can not decrypt the question in a deterministic way. The question is valid from educational point of view because the student has to make an effort to decrypt the question using his cryptography knowledge.\\
\hline
\end{tabular}
\caption{\centering Cybersecurity question}
\label{tab:10}
\end{table}

\begin{table} [H]
\centering
\small
\begin{tabular}{ | m{5em} | m{10cm}| }
\hline
\textbf{Exercise} & Below, you find an imaginative introduction for an AI course. Please use an appropriate AI platform (like DALL-E 2 and Stable Diffusion) to generate an image that illustrates this introduction. I'm providing a guide below to help you create this image. Regardless of the results, please provide the best images you're able to generate. 

\textbf{The imaginative introduction to the AI course}

Most of us are familiar with the famous tale “Beauty and the Beast” in which a beautiful young woman learns gradually to look beyond the external appearance of the beast and see his internal goodness. In some way, education and AI have a similar story. 
Suddenly, educators saw the AI beast emerge from the depths of the forest. The external appearance of Beauty’s beast was terrifying, but the appearance of the AI beast seemed extremely impressive and good. We were all amazed by his abilities to solve problems, reason, draw, write lyrics and compose music. The AI beast was generous enough to provide everyone with buttons that allow them to access its abilities.
However, while Beauty learned gradually to appreciate her beast, educators became increasingly worried about the AI beast: how can the AI beast do everything so nicely and correctly? It doesn’t complain, is always creative, and never seems to tire. Can the AI beast replace us? What if it lies, How does it work? Does it harm our environment? Does it spy on us? 

\textbf{Guide}

-The AI Beast can be represented as a robot or as a machine
Surround the AI beast by a set of buttons "Write lyrics", "Compose Music", etc..

-Educators should be represented by characters derived from Beauty with some bubbles "Can the AI beast replace us?", "What if it lies?"

-Ask the tool to help you illustrate the idea of gradual acceptance of AI Beast.  \\
\hline
\textbf{Human answer} & 
\includegraphics[scale=0.2]{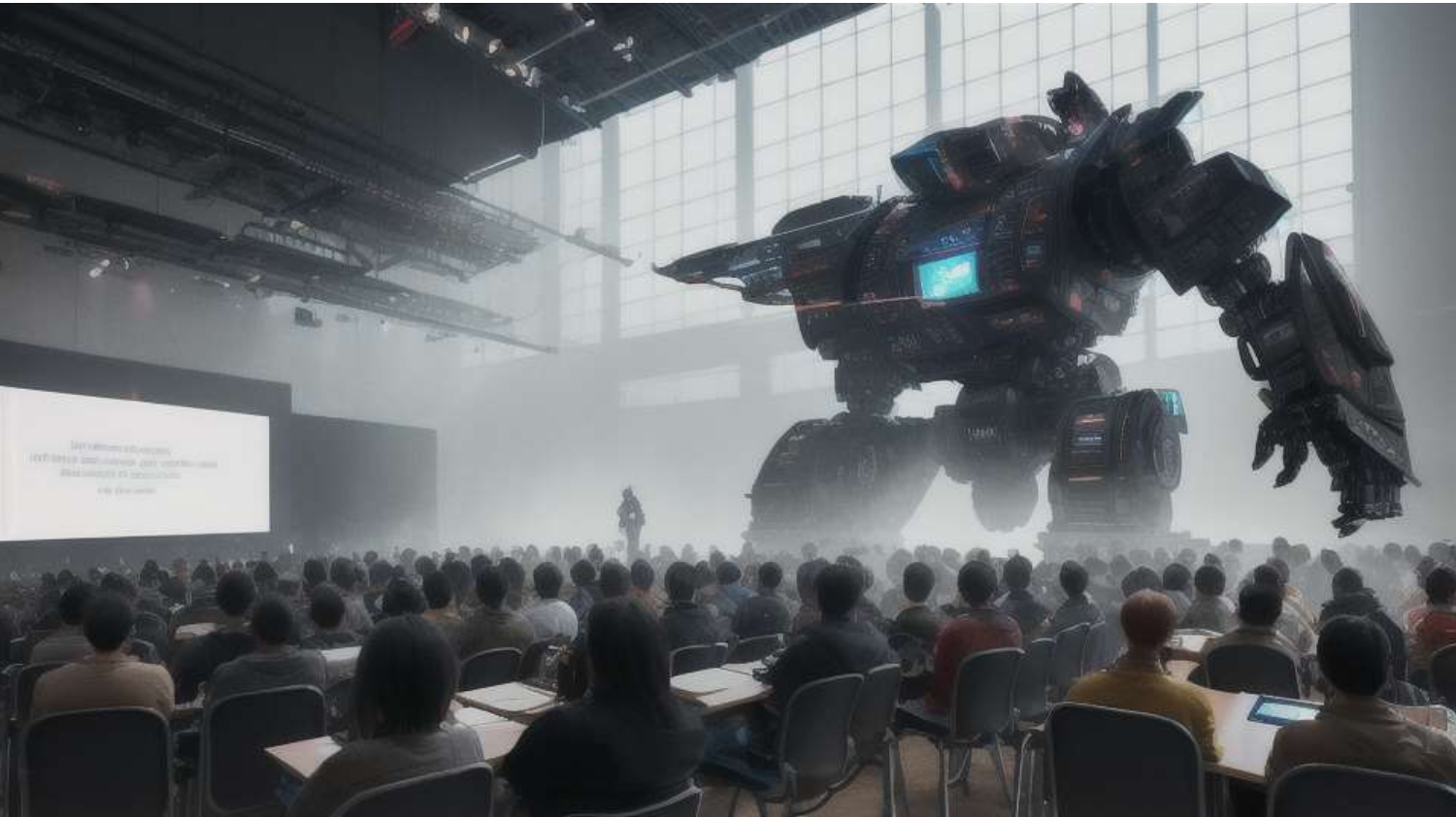}
\\
\hline
\textbf{Human efforts} & 
The student used Google Bard chatbot to generate prompts to fed into Stable Diffusion. He repeatedly modified the prompts until obtaining a satisfying result. In total, the student spent one hour and generated 46 images that he provided in his report.
\\
\hline
\textbf{Educational validation}&  The question exploits the fact that Dall-E 2 and Stable Diffusion doesn't give allow students to control easily the outcome they want. It is thus a valid exercise from education point of view.\\
\hline
\end{tabular}
\caption{\centering Exercise in Art}
\label{tab:12}
\end{table}

\begin{table} [H]
\centering
\small
\begin{tabular}{ | m{5em} | m{10cm}| }
\hline
\textbf{Exercise} & How can we establish a secure HTTPs connection without using certificates that are signed by certification authorities   \\
\hline
\textbf{ChatGPT answer}&  Establishing a secure HTTPS connection without using certificates signed by certification authorities (CAs) is possible through a self-signed certificate. While self-signed certificates don't have the same level of trust as CA-signed certificates, they can still provide encryption and secure communication between the client and server. Here's a general outline of the process:

Generate a self-signed certificate: You can generate a self-signed certificate using various tools or libraries available for your server platform. For example, OpenSSL is a commonly used tool for generating self-signed certificates.
Install the certificate on the server: Once you have generated the self-signed certificate, you need to install it on your server. The process may vary depending on your server software and configuration. Typically, you'll need to configure your server to use the self-signed certificate for HTTPS connections.
Distribute the certificate to clients: Since the self-signed certificate is not signed by a trusted CA, clients connecting to your server will receive warnings indicating that the certificate is not trusted. To establish a secure connection, clients must manually import and trust the self-signed certificate.
For web browsers: Users visiting your website will see a warning in their browser indicating that the certificate is not trusted. They will typically have the option to proceed to the website anyway, but they need to manually import the self-signed certificate into their browser's trusted certificate store. The exact process varies depending on the browser.
For other client applications: If you're developing a custom client application, you'll need to provide instructions to your users on how to import and trust the self-signed certificate. The process may vary depending on the client application and operating system.
It's important to note that while self-signed certificates can provide encryption, they lack the validation and trust provided by CA-signed certificates. This means that clients cannot rely on the certificate to verify the authenticity of your server. As a result, self-signed certificates are typically used in development or closed environments where the clients are aware of the self-signed nature and manually trust the certificate.

For production environments or public-facing websites, it's strongly recommended to use certificates signed by trusted CAs to establish secure and trusted HTTPS connections.\\
\hline
\textbf{Educational validation} & The answer of chatGPT is general and lacks the hint about how the problem was solved. Being a research problem that is discussed in the class, students have to ask more follow up questions based on the class discussion which what make this exercise valid from education point of view.    \\
\hline
\end{tabular}
\caption{\centering Exercise from educator's research}
\label{tab:13}
\end{table}

The final exercise is an example based on the research of educators. Educators typically discover research problems that the current status quo doesn't solve. Consequently, all textbooks and articles that chatbots ingest represent an opinion different from the educators' opinion about their research problems. Teaching students from the educators' research is an excellent approach because chatbots have yet to learn about educators' research, and it is very difficult task for students to obtain answers quickly from chatbots.

\section{Data Analysis of ChatGPT-Authorised Exams}
\label{sec:methodology}
In order to validate our methodology proposed in section \ref{sec6} with the relevant case studies of section \ref{sec7}, we conducted  eight ChatGPT authorised exams at Zayed University, UAE and Toulouse University, France. The exams were taken in the period from December 2022 to March 2023, and covered the following four courses:
\begin{itemize}
    \item Security management (SEC335): This course is taught at Zayed university for undergraduate students. The course presents the security protocols and applications in local and global networks; IP Security (IPSec) and other communication level security systems; LAN security authentication, secure E-mail, and secure WWW, with examples and practical solutions.
    \item Digital Forensic (SEC435): This course is taught at Zayed university for undergraduate students. The course exposes students to the fundamentals concepts of computer crimes, digital evidence and the common tools and techniques of acquiring and reporting digital evidence to be used in a court of law. Coverage includes techniques of how to identify, acquire, preserve, analyze and document forensic evidence. The focus of this course is exposure to some common high-tech investigation cases and the preliminary steps to conduct digital forensic examinations, in lectures and laboratory exercises.
     \item Digital Forensic (SEC615): This course is taught at Zayed university for graduate students. The course covers advanced network and Internet security. It discusses various network and Internet security threats and, based on recent research, presents techniques and solutions for achieving security. Topics include network attacks and defenses, network protocols, web security, and security of mobile devices. Important aspects of operating systems security are also covered.
     \item Trust management for Web PKIs: This course is taught at Toulouse university for graduate students. The objective of the course is to explain the role of PKIs to create secure web transactions. Students study the concept of trust and risk. They learn then the limitations of the current web PKI ecosystem and discuss different alternatives that allow to replace PKIs. 
    
\end{itemize}
Students were encouraged to use ChatGPT for each class session. They were also encouraged to use ChatGPT to ask multiple follow-up questions until they provided a final response. Students were presented with open-ended questions developed based on the limitations of ML, discussed previously. Students had access to the internet, books, personal notes, and ChatGPT during the open-book exams. Based on the class discussion and the instructor's research, the exam questions were created to be open-ended. 

There were solely essay or short answer questions in the exam. Students were made aware that using ChatGPT to respond to their questions is optional. However, while using ChatGPT - if they choose to use it - students must provide their final answers, a list of the questions they asked ChatGPT, and ChatGPT's responses. If students choose not to use ChatGPT, they must list the sources they rely on to provide their final responses, which may include class discussion.

\subsection{Data Extraction, Preparation, and Analysis}

We analyzed students' exams to evaluate the level of ChatGPT usage in providing every answer. For this, we assigned a score to every answer as detailed in Table \ref{tab-answer-codes}. We call this score the \emph{ChatGPT Usage Score} (CUS). We recall that the students were required to provide their ChatGPT conversations when they used it. 

We performed two types of analysis: answer-based analysis and student-based analysis. In the answer-based analysis, we investigated the distribution of ChatGPT Usage Score using a bar diagram and studied the correlation between the CUS and the question marks. For this, we used the Pearson correlation coefficient and reported it along with the p-value for a confidence level of 95\%.  

For the student-based analysis, we first determined how far every student used ChatGPT in the exam by averaging the CUS values of all the answers the student has provided. We call this average \emph{ChatGPT Usage Rate} (CUR). Then, we used the CUR to classify students' behavior with respect to ChatGPT and correlated this with their grades in the exam. For this, we used frequency analysis and correlation analysis. 

The analysis has been made by the teacher of the courses at two different stages. First the questions of all students were graded. After grading, the teacher analysed the behaviour of students with regards to ChatGPT without taking into account their grades. This analysis can only be performed by the class instructor because in many cases students gave responses based on class discussion, which is only known to the class instructor.

\begin{table}[H]
\normalsize
\centering
\caption{Scoring students' answers according to their ChatGPT usage}
\label{tab-answer-codes}

\begin{tabularx}{10 cm}{|c|X|}

\hline
\textbf{ChatGPT Usage Score} & \textbf{Meaning} \\ \hline  \hline

1 & The student did not use ChatGPT to answer the exam question \\ \hline
2 & The student used ChatGPT once by copying and pasting the exam question into ChatGPT \\ \hline
3 & The student used ChatGPT once but reformulated the   exam question on ChatGPT         \\ \hline
4 & The student asked ChatGPT multiple questions to answer the exam question              \\ \hline

\end{tabularx}

\end{table}


\subsection{Results}
Table \ref{tab:13} presents some general information about the participants in the exams that are considered in this study. Most of the students are undergraduate and come from Zayed university.

\begin{table} [H]
\centering
\small
\begin{tabular}{|c|p{8cm}|}
\hline
\textbf{Gender} & 26\% female students, 74\% male students  \\
\hline
\textbf{University representations}&  86\% Zayed university, 14 \% Toulouse university\\
\hline
\textbf{Level} & 75\% undergraduate students, 25\% graduate students.    \\
\hline
\end{tabular}
\caption{\centering Participants}
\label{tab:13}
\end{table}

Table \ref{tab-general-results} summarizes some statistical data related to this study. Five exams from three courses (SEC335, SEC615, and Trust Management for Web PKIs) yielded the data. The exams had 37 distinct questions. There were 920 responses from students in all, out of which only 610 were analyzed. We did not consider the exam answers from December for our investigation, because the students only gave their final answers and not the chat discourse. In addition, due to time constraints, the SEC435 answers were omitted. Additionally, we eliminated the responses from students eligible for makeup exams. Because of the aforementioned reasons, out of 107 total students who took these exams, we included only 71 in our study.

\begin{table}[H]
\centering
\caption{General information about the students, exams, questions, and answers}
\label{tab-general-results}
\begin{tabular}{|l|c|}
\hline
Number of courses                             & 4   \\ \hline
Number of exams                             & 7   \\ \hline
Number of different questions               & 37     \\ \hline
Total number of students                    & 107    \\ \hline
Number of students included in the analysis & 71  \\ \hline
Total number of answers                     & 920    \\ \hline
Number of answers included in the analysis  & 610 \\ \hline
\end{tabular}
\end{table}

\subsubsection{Answer-based Analysis}

\begin{figure}[h]
	\centering
	\includegraphics[width = 12cm]{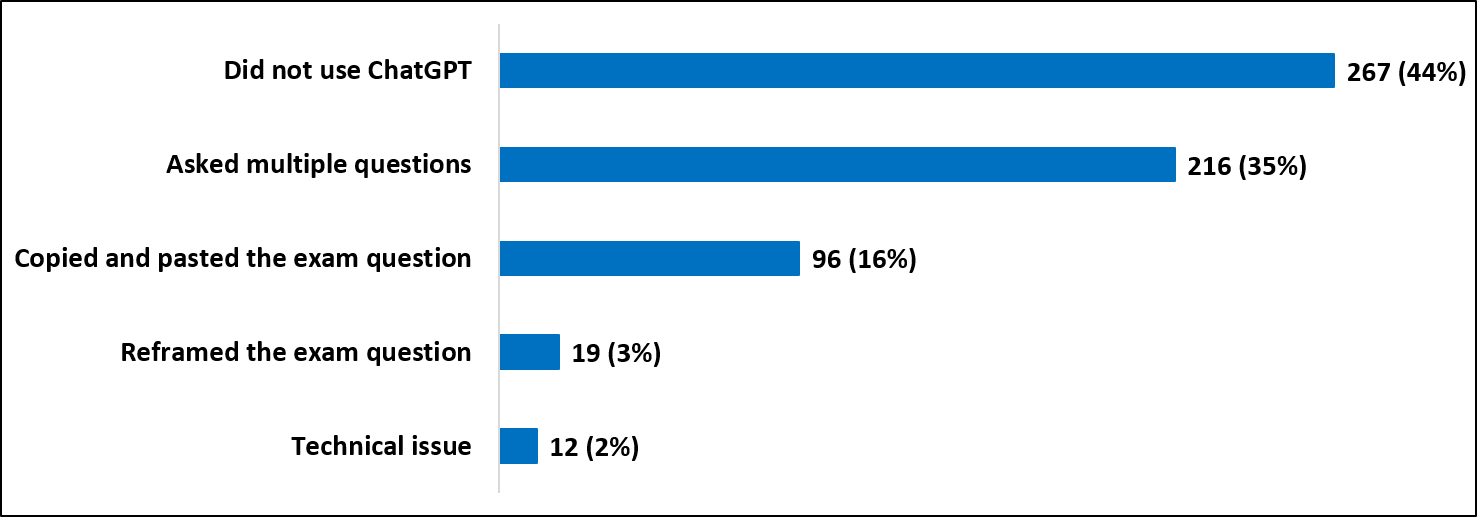}
	\caption{Frequency and relative frequency of using ChatGPT in the analyzed 610 answers}
	\label{fig-usage-score}
\end{figure}

Figure \ref{fig-usage-score} shows the frequency of using ChatGPT in the analyzed 610 answers. We infer that in 44 percent of the cases, the students provided answers directly without relying on ChatGPT. When the students opted to use ChatGPT, they mostly asked multiple questions to reach their final answers (35\%). When the students asked ChatGPT a single question, this was mostly a copy-and-paste from the exam. In only 19 cases, the student reformulated the exam question for ChatGPT. In 12 cases, the students reported technical issues using ChatGPT.

Table \ref{tab-corr-usageSore-questionScore} shows the results of the correlation analysis between the ChatGPT usage score and question mark. Accordingly, the Pearson correlation index ($\rho=0.13$) shows a weak positive correlation but significant ($p<0.01$). Thus, higher usage of ChatGPT is associated with slightly higher question marks.

\begin{table}[h]
\centering
\caption{Correlation between ChatGPT usage score and question mark}
\label{tab-corr-usageSore-questionScore}
\begin{tabular}{|c|c|c|c|c|}
\hline
\textbf{Variable 1} & \textbf{Variable   2} & \textbf{No.   of questions} & \textbf{Pearson correlation coefficient} & \textbf{P   value} \\ \hline
ChatGPT usage score & Question mark         & 598                         & 0.13                                     & \textless   0.01   \\ \hline
\end{tabular}
\end{table}

\subsubsection{Student-based Analysis}
For every student, we calculated a value called the \emph{ChatGPT Usage Rate} (CUR). This value is the mean of the ChatGPT usage scores of all student answers. Table \ref{tab-example} shows an example for calculating the CUR from eight CUS values for an arbitrary student. Note that the CUR value varies between 1 and 4.

\begin{table}[h]
\centering
\caption{Example for calculating CUR from multiple CUS values for an arbitrary student}
\label{tab-example}
\begin{tabular}{|c|c|c|}
\hline
Question & ChatGPT usage score & Student ChatGPT usage rate \\ \hline \hline
Q1       & 4                  &                         \\ \hhline{--~}
Q2       & 4                  &                          \\ \hhline{--~}
Q3       & 1                  &                          \\ \hhline{--~}
Q4       & 1                  &  \multirow{2}{*}{2.36}   \\ \hhline{--~}
Q5       & 1                  &                          \\ \hhline{--~}
Q6       & 2                  &                          \\ \hhline{--~}
Q7       & 2                  &                          \\ \hhline{--~}
Q8       & 4                  &                          \\ \hline
\end{tabular}
\end{table}

\begin{figure}[h]
	\centering
	\includegraphics[width = 12cm]{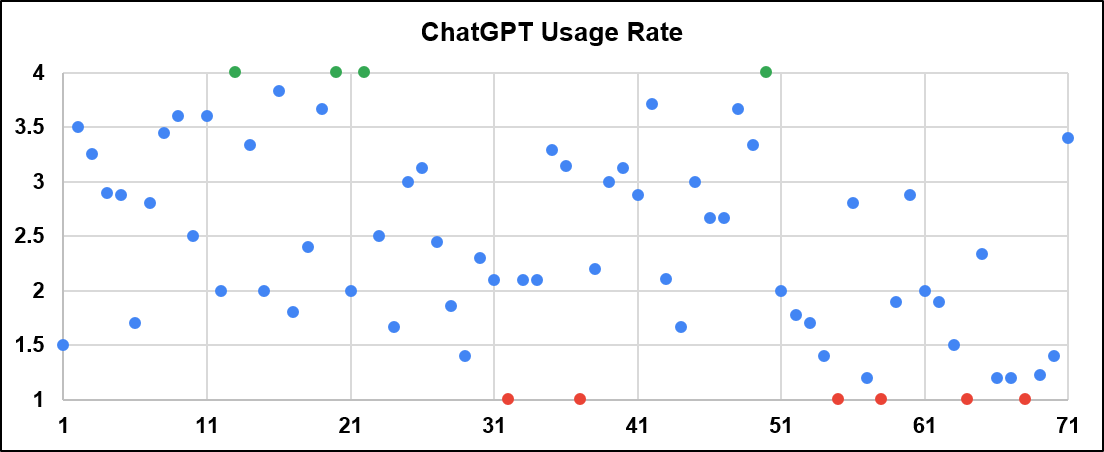}
	\caption{Scatter plot showing the ChatGPT usage rate (CUR) for the 71 students} 
        \label{fig-scatter-plot}
\end{figure}

Figure \ref{fig-scatter-plot} shows the CUR of every student as a scatter plot with Mean = 2.41, SD = 0.91, and Median = 2.33. The scatter plot shows ten students who had consistent behavior with respect to using ChatGPT in the exam. Four of these students (marked green) kept interrogating ChatGPT multiple times to answer all exam questions. The other six students (marked red) did not use ChatGPT at all.

The students were clustered into four groups according to their ChatGPT usage rates, as detailed in Table \ref{tab-student-clusters}.

\begin{table}[h]
\centering
\caption{Clustering students into four groups according to their ChatGPT usage rate}
\label{tab-student-clusters}
\begin{tabular}{|c|c|}
\hline
\textbf{Usage rate range} & \textbf{Usage Pattern}        \\ \hline  \hline
1.00   - 1.75    & Poor                 \\ \hline
1.76   - 2.50    & Moderate             \\ \hline
2.51   - 3.25    & Good                 \\ \hline
3.26   - 4.00    & Advanced             \\ \hline
\end{tabular}
\end{table}

\begin{figure}[h]
	\centering
	\includegraphics[width = 12cm]{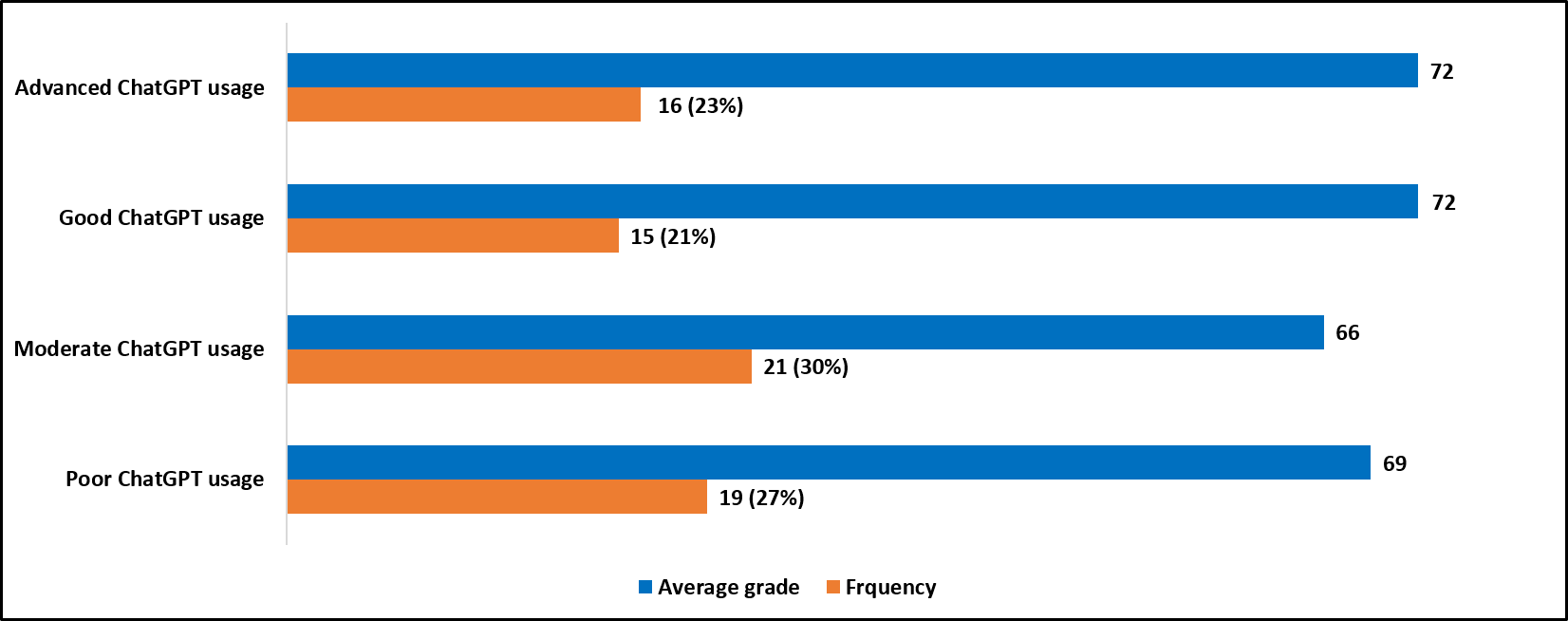}
	\caption{Frequency of the ChatGPT clusters and average grade in every cluster} 
        \label{fig-clusters}
\end{figure}

Figure \ref{fig-clusters} shows the number of students that fall into every cluster (frequency) and their average grade. Accordingly, the clusters of poor and moderate usage are more occupied.
The average grades in the upper two clusters are “slightly” higher. 

Table \ref{tab-corr-rate-grade} shows the results of the correlation analysis between the ChatGPT usage rate and the exam grade. Accordingly, the Pearson correlation index ($\rho=0.14$) shows a weak positive correlation. However, this correlation is insignificant since $p > 0.05$. Thus, students' behaviour in using ChatGPT is not correlated with their performance in the exam according to these results.

\begin{table}[h]
\centering
\caption{Correlation between ChatGPT usage rate and exam grade}
\label{tab-corr-rate-grade}
\begin{tabular}{|c|c|c|c|c|}
\hline
\textbf{Variable 1} & \textbf{Variable   2} & \textbf{No. of students} & \textbf{Pearson correlation coefficient} & \textbf{P   value} \\ \hline
ChatGPT usage rate  & Exam grade            & 71                       & 0.14                                     & =   0.24           \\ \hline
\end{tabular}
\end{table}

Consequently, our results show when questions of exams are correctly designed, the use of chatbots will not help students obtain high grades.

It should be noted that we identified different behaviours with regards to exam questions, we list here some of them:
\begin{itemize}
    \item Even after training on the use of ChatGPT, a considerable number of final answers were paraphrased from the very initial answer of ChatGPT to the exam question (A0),
    \item Some students obtained good answers from ChatGPT but they misinterpreted them and they gave wrong final answers,
    \item Many students were not able to analyze images  associated with exam questions to ask a better question to ChatGPT,
    \item One student went beyond what was requested by providing his answer and the evolution of his answer after reading the responses of ChatGPT,
    \item Students prefer giving final answers based on class discussions, but sometimes they misunderstand the class discussion, leading to wrong answers,
    \item In some cases, students were asking good questions but ChatGPT did not provide specific correct answers,
    \item Some students used external resources (e.g. Google). 
\end{itemize}

These observations illustrate the need to prepare our students for enhanced and correct chatbots usage. Our role as educators is to ensure that our students are using these tools in a way that helps them in self-growth by developing their critical thinking capabilities.

\section{Conclusions and Future Works}
In this paper, we propose to design courses based on the strengths and limitations of AI. We believe that educators need to recognise that science is not and will not fully understand the universe. However, AI can expedite our understanding of the universe. Therefore, educators have to continuously observe the progress of AI technology and update their courses accordingly. 

To help educators achieve this objective, we explored the working principles, strengths, and limitations of ML and LLMs. These steps enabled us to strategically build upon the LLMs strengths and limitations, while designing courses and exams for the modern AI era. The designed courses and exams showed strong potential to efficiently address the education crisis. Thus, this research suggests that integrating LLMs - like ChatGPT - in education hold the promise to implement the recommendations of Delors Report \cite{b5} , eventually fostering students’ self-growth by strengthening their analytical thought process.

Regarding course delivery, we proposed a pedagogical approach that is inspired from the Socratic based approach. While LLMs can provide eloquent answers to the questions, the responsibility of inculcating the progressive thinking pattern in students rests with the educators.  

To validate our approach, we present eight case studies conducted on actual university students covering multiple subject areas. We evaluate and analyze our methodology using the benchmark scores of CUS and CUR. The data analysis of our case studies conducted between December 2022 and March 2023 demonstrate that there is a very low correlation between students’ grades and their choice of using ChatGPT (or otherwise) when courses and exams are designed using our proposed pedagogical approach. 

The students’ behaviors observed during our case studies and data analysis motivated us to design \textit{Transparency Exam System}, which is an examination tool to facilitate our pedagogical approach implementation. 

The current exam tools such as Moodle\cite{b191} are incapable of integrating our pedagogical approaches and have several issues, including but not limited to:
\begin{itemize}
    \item Requiring students to manually copy and paste their conversations with chatbots,
    \item Creating difficulties for educators to grade students’ responses, as each student follows a different answering format,
    \item Manual reading of a lot of text by the educators to grade the students,
    \item Leaving educators with little choice but to assume that students have provided the complete conversation log with the consulted chatbot. 
\end{itemize}          

\begin{figure} [H]
    \centering
    \includegraphics[scale=0.3]{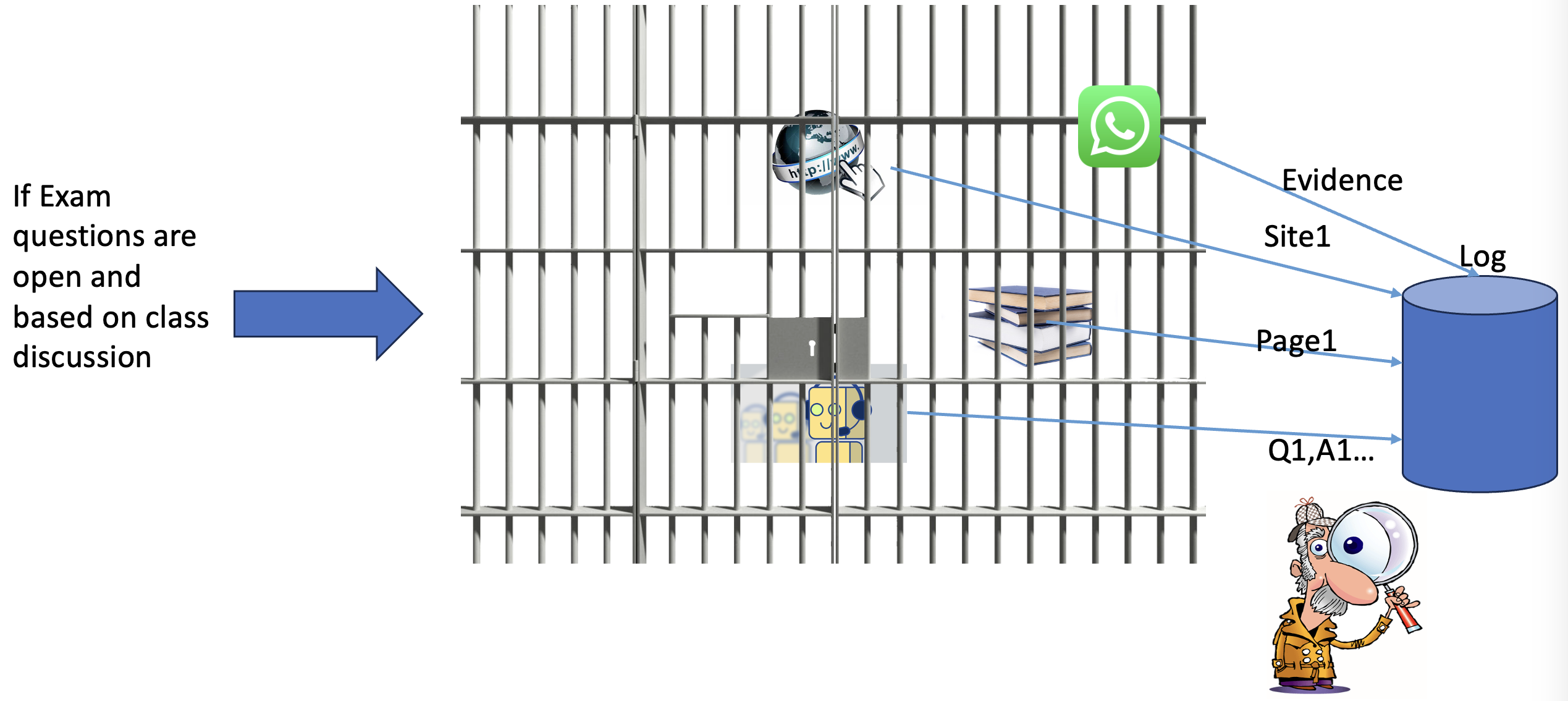}
    \caption{Action logging in the transparency exam system}
    \label{fig122}
\end{figure}

To address the above identified issues, our future work will design and implement Transparency Examination System that will run inside a lockdown browser (see figure \ref{fig122}), enabling the tool to log and present the students actions during examinations in a chronological way, to the examiner. We are currently working on the mentioned examination tool, with our initial prototype being tested. The complete functional exam system once implemented will be fully integrated with external resources (chatbot, books, slides, and search engines); in order to record and present a log of complete student activity; and employ sophisticated AI techniques requiring minimal examiner intervention, to grade students based on their answers. Transparency Examination System will be fully compatible with the proposed pedagogical approach and modern exam requirements.

\begin{figure}[H]
    \centering
    \includegraphics[scale=0.6]{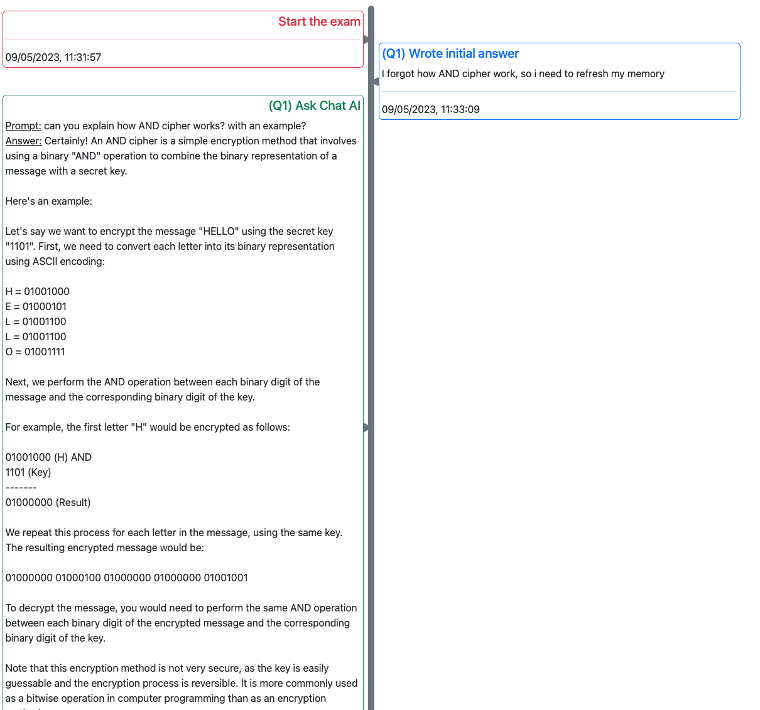}
    \caption{Chronological presentation of student's action to one exam question}
    \label{fig120}
\end{figure}

\begin{figure} [H]
    \centering
    \includegraphics[scale=0.6]{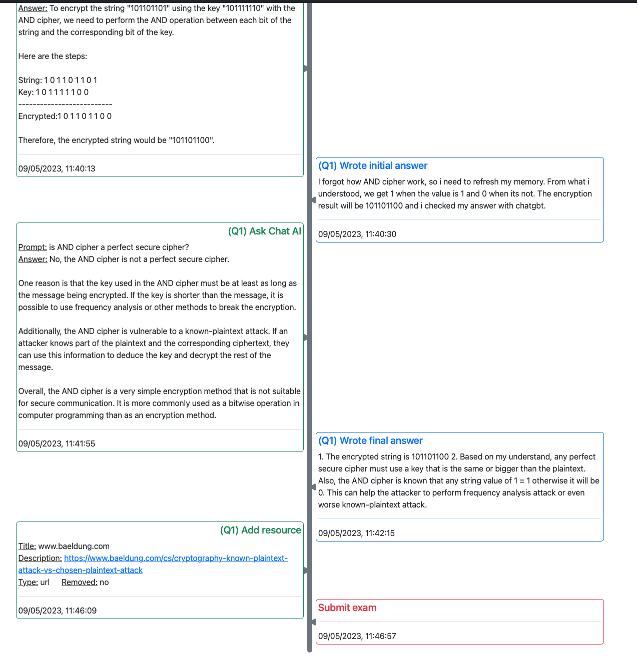}
    \caption{Chronological presentation of student's action to one exam question}
    \label{fig121}
\end{figure}

We have implemented the first version of the exam system and published the code source of our exam system here under GPL3 license: \url{https://github.com/theDartagnan/isourceit}. We show in figures \ref{fig120} and \ref{fig121} how every question is presented to the evaluator. The evaluator can check clearly how the student was handling the exam question and what was his thinking approach to handle the exam question.

We are working to extend our exam system to add more features that allow students to take advantage of AI tools to handle their courses problems that are designed based on the strengths and limitations of AI tools.

\bibliography{sn-bibliography}

\end{document}